\documentclass[aps,prb,twocolumn,superscriptaddress,showpacs]{revtex4-2}
\usepackage{graphicx}%
\usepackage{amssymb}
\usepackage{amsmath}
\usepackage{color}

\newcommand{\ET}{(BEDT\--TTF)$_2$$X$}

\newcommand{\CuCN}{$\kappa$-(BEDT\--TTF)$_2$\-Cu$_2$(CN)$_3$}
\newcommand{\AgCN}{$\kappa$-(BEDT\--TTF)$_2$\-Ag$_2$(CN)$_3$}
\newcommand{\CuCl}{$\kappa$-(BEDT\--TTF)$_2$\-Cu[N(CN)$_2$]Cl}

\begin{document}
\preprint{k-CuCN dielectrics pressure}

\title{Phase coexistence at the first-order Mott-transition revealed by \\
pressure-dependent dielectric spectroscopy of $\kappa$-(BEDT\--TTF)$_2$\-Cu$_2$(CN)$_3$}

\author{R. R\"osslhuber}
\affiliation{1.~Physikalisches Institut, Universit\"{a}t
	Stuttgart, Pfaffenwaldring 57, 70569 Stuttgart, Germany}
\author{A. Pustogow}
\affiliation{1.~Physikalisches Institut, Universit\"{a}t
	Stuttgart, Pfaffenwaldring 57, 70569 Stuttgart, Germany}
\affiliation{Department of Physics and Astronomy, UCLA, Los Angeles, California 90095, U.S.A.}
\affiliation{Institute of Solid State Physics, Vienna University of Technology, 1040 Vienna, Austria.}
%\altaffiliation{Present Address: Institute of Solid State Physics, Vienna University of Technology, 1040 Vienna, Austria.}
\author{E. Uykur}
\affiliation{1.~Physikalisches Institut, Universit\"{a}t
	Stuttgart, Pfaffenwaldring 57, 70569 Stuttgart, Germany}
\author{A. B\"ohme}
\affiliation{1.~Physikalisches Institut, Universit\"{a}t
	Stuttgart, Pfaffenwaldring 57, 70569 Stuttgart, Germany}
\author{A. L\"ohle}
\affiliation{1.~Physikalisches Institut, Universit\"{a}t
	Stuttgart, Pfaffenwaldring 57, 70569 Stuttgart, Germany}
\author{R.~H\"ubner}
\affiliation{Institute for Functional Matter and Quantum Technology, Universit\"{a}t
	Stuttgart, 70569 Stuttgart, Germany}
\author{J. A. Schlueter}
\affiliation{Material Science Division, Argonne National Laboratory,
Argonne, Illinois 60439-4831 and \\ National Science Foundation, Alexandria, Virginia 2223, U.S.A.}
\author{Y. Tan}
\affiliation{National High Magnetic Field Laboratory, Florida State University, Tallahassee, U.S.A.}
\author{V. Dobrosavljevi\'c}
%\email{vlad@magnet.fsu.edu}
\affiliation{National High Magnetic Field Laboratory, Florida State University, Tallahassee, U.S.A.}
\author{M. Dressel}
%\email{dressel@pi1.physik.uni-stuttgart.de}
%\homepage{http://www.pi1.uni-stuttgart.de}
\affiliation{1.~Physikalisches Institut, Universit\"{a}t
	Stuttgart, Pfaffenwaldring 57, 70569 Stuttgart, Germany}
\date{\today}

\begin{abstract}
The dimer Mott insulator $\kappa$-(BEDT-TTF)$_2$Cu$_2$(CN)$_3$ can be tuned into metallic and superconducting states upon applying pressure of 1.5~kbar and more.
We have performed dielectric measurements (7.5~kHz to 5~MHz) on $\kappa$-(BEDT-TTF)$_2$Cu$_2$(CN)$_3$ single crystals as a function of temperature (down to $T=8$~K) and pressure (up to $p=4.3$~kbar).
In addition to the relaxor-like dielectric behavior seen below 50~K at $p=0$, that moves toward lower temperatures with pressure,  a second peak emerges in $\varepsilon_{1}(T)$ around $T=15$~K.
When approaching the insulator-metal boundary, this peak diverges rapidly reaching $\varepsilon_{1} \approx 10^{5}$. Our dynamical mean-field theory calculations substantiate that the dielectric catastrophe at the Mott transition is not caused by closing the energy gap, but due to the spatial coexistence of correlated metallic and insulating regions. We discuss the percolative nature of the first-order Mott insulator-to-metal transition in all details.

\end{abstract}

\date{\today}
\pacs{
71.30.+h,  % Metal-insulator transitions and other electronic transitions
74.70.Kn,  % Organic superconductors
%78.30.Jw,  %  Organic compounds, polymers
72.90.+y,  % Other topics in electronic transport in condensed matter
77.22.-d    %Dielectric properties of solids and liquids
}

\maketitle

\section{Introduction and Motivation}
\label{sec:Introduction}
The quasi two-dimensional organic charge-transfer salts \ET\ became model compounds for investigating the interplay of strongly correlated electrons, reduced dimensionality, spin-charge interactions and ordering phenomena \cite{Seo04,LebedBook,Powell11,Dressel2020}. Their molecular composition enables fine-tuning of physical properties by modifying the donor molecules bis-ethylene\-dithio-tetra\-thia\-ful\-valene (BEDT-TTF), varying the monovalent anions $X$, or by applying comparably low pressures -- usually a few kbar induce drastic changes \cite{IshiguroBook,*ToyotaBook,*MoriBook}. The dimerized $\kappa$-phase compounds have been established as prime examples for the bandwidth-tuned Mott insulator-metal transition (IMT) and the quantum critical region above \cite{Lefebvre00,Limelette03a,Kagawa05,*Kagawa09,Furukawa15,*Furukawa18,Pustogow18b,Terletska11,
*Vucicevic15}.

\begin{figure}[h]
	\centering
	\includegraphics[width=0.6\columnwidth]{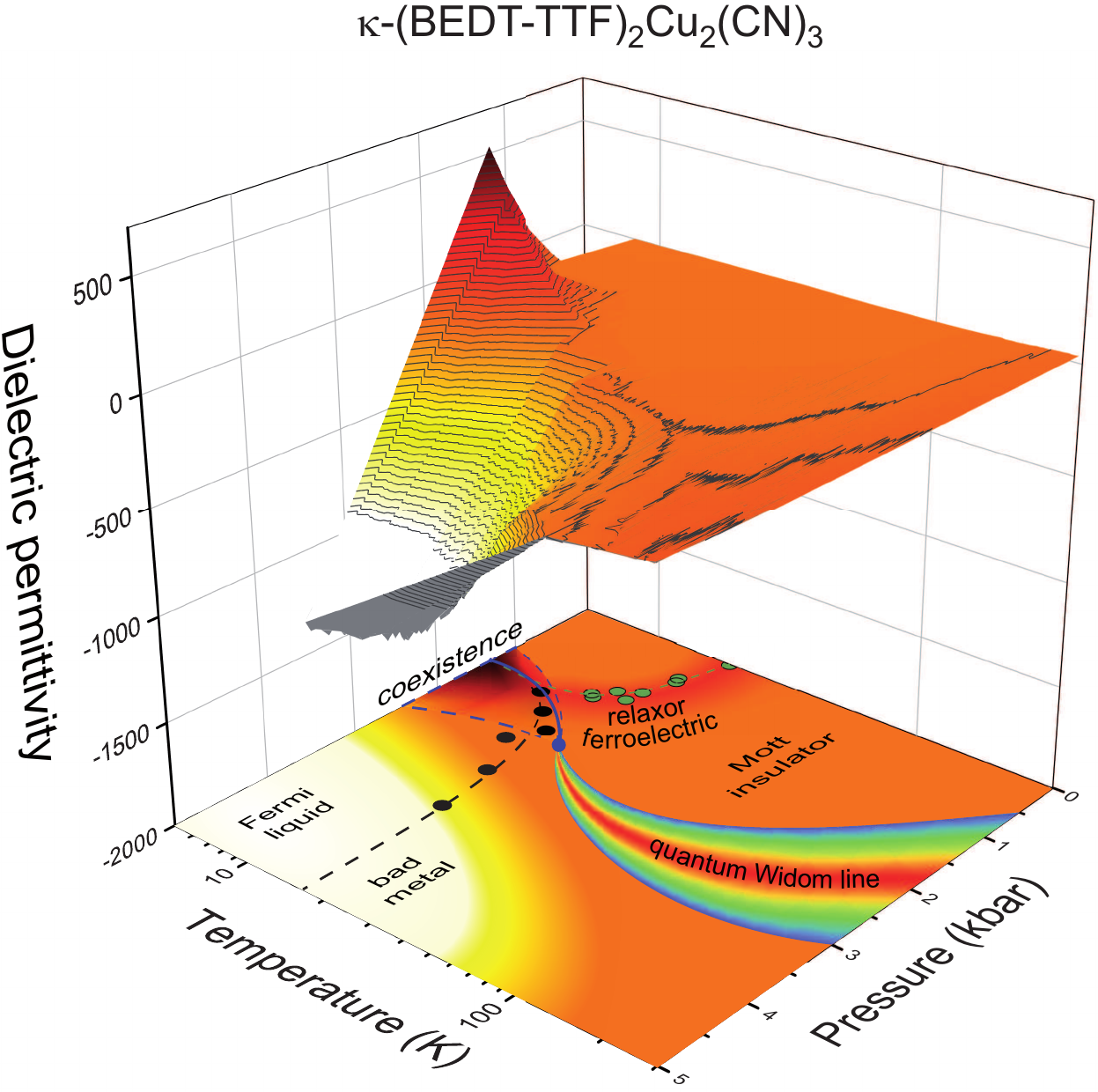}
	\caption{3D plot of $\varepsilon_{1}(p,T,f=380$~kHz) and phase diagram of \CuCN.
	Strikingly, the dielectric permittivity  is strongly enhanced at the first-order Mott transition (blue line around 1.8~kbar and $T<T_{\rm crit}\simeq 16$~K), ascribed to metal-insulator phase coexistence, as predicted by dynamical mean-field theory \cite{Vucicevic13}. Above $T_{\rm crit}$, the first-order IMT becomes a gradual crossover (quantum Widom line) \cite{Geiser91,Furukawa15,Pustogow18b}. A relaxor-ferroelectric response in $\varepsilon_{1}(T)$ is observed in the Mott-insulating phase ($p<1.5$~kbar) \cite{Abdel10,Pinteric14,*Pinteric15,*Pinteric18}. The green circles represent the bifurcation temperature $T_{\rm B}$ that indicates a change in the relaxation mechanism as discussed in Sec.~\ref{sec:Mode1}. The black solid circles correspond to $T_{\rm FL}$ \cite{Pustogow2021-percolation}.
	}
	\label{Phase_diagram}
\end{figure}

Since two decades, the dimer Mott insulator \CuCN\ attracted most attention
as it has been considered the prime candidate for a quantum spin liquid  \cite{Shimizu03,*Kurosaki05,Kanoda11}.
The nature of the low-temperature spin state remains subject of intense studies and controversial discussions \cite{Savary17,Zhou17}, fueled by the recent finding of a spin gap~\cite{Miksch2020}.
The absence of long-range magnetic order even at lowest temperatures
provides the opportunity to investigate the $genuine$ Mott transition  \cite{Furukawa18,Dressel18,*Pustogow18a,Pustogow18b}, which is solely driven by Coulomb interactions, without breaking any symmetry. Nevertheless, recent studies revealed that for these layered BEDT-TTF compounds the lattice properties play an important role --~in addition to disorder~-- and the interaction with the anions can be decisive \cite{Dressel16,Pouget18,Foury-Leylekian18}.

\begin{figure}
	\centering
	\includegraphics[width=1.0\columnwidth]{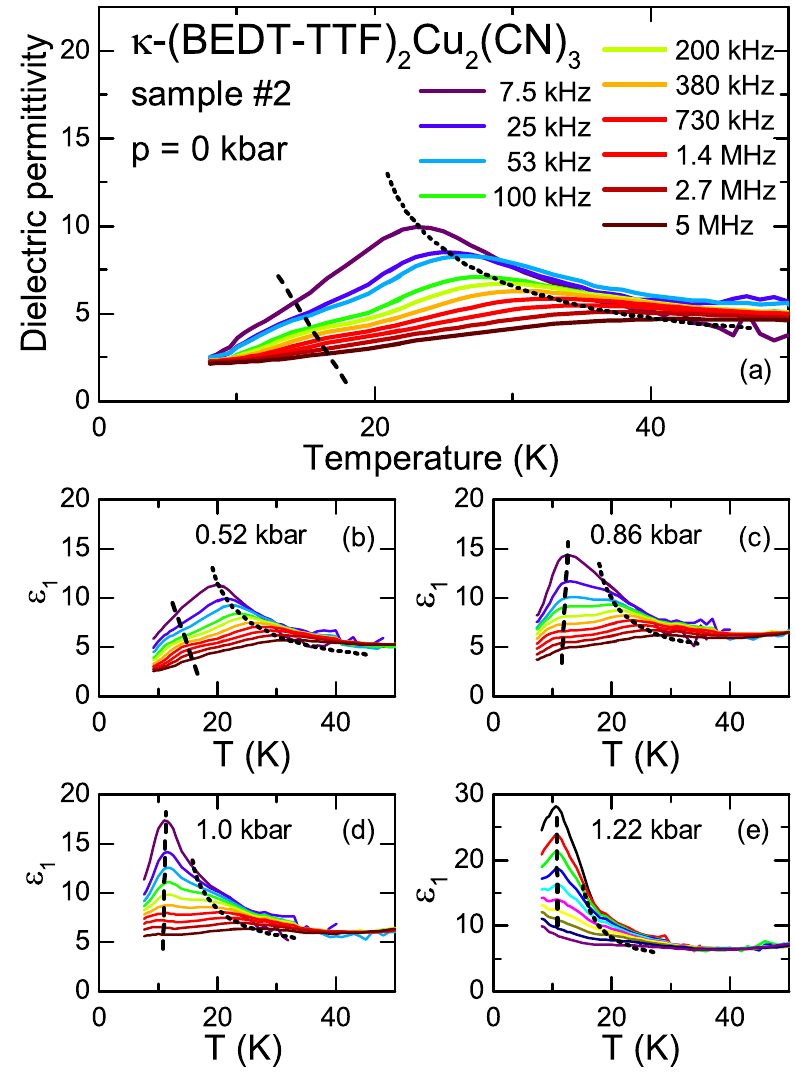}
	\caption{Temperature-dependent dielectric constant $\varepsilon_{1}(T)$ of \CuCN\ at low pressure values ($p=0$--1.22~kbar)  plotted for frequencies 7.5~kHz--5~MHz. (a)~At ambient pressure and $T<50$~K, the dielectric permittivity exhibits a relaxor-type ferroelectric behavior with a peak that diminishes in amplitude and shifts to higher $T$ as frequency increases.
(b-e)~In addition, we identify another, shoulder-like feature at lower temperature. While traces are revealed around $T=15$~K already at $p=0$, it forms a second peak with distinct temperature dependence with increasing pressure. Note the different ordinates.}
	\label{eps(T)}
\end{figure}

\begin{figure}
	\centering
	\includegraphics[width=1.0\columnwidth]{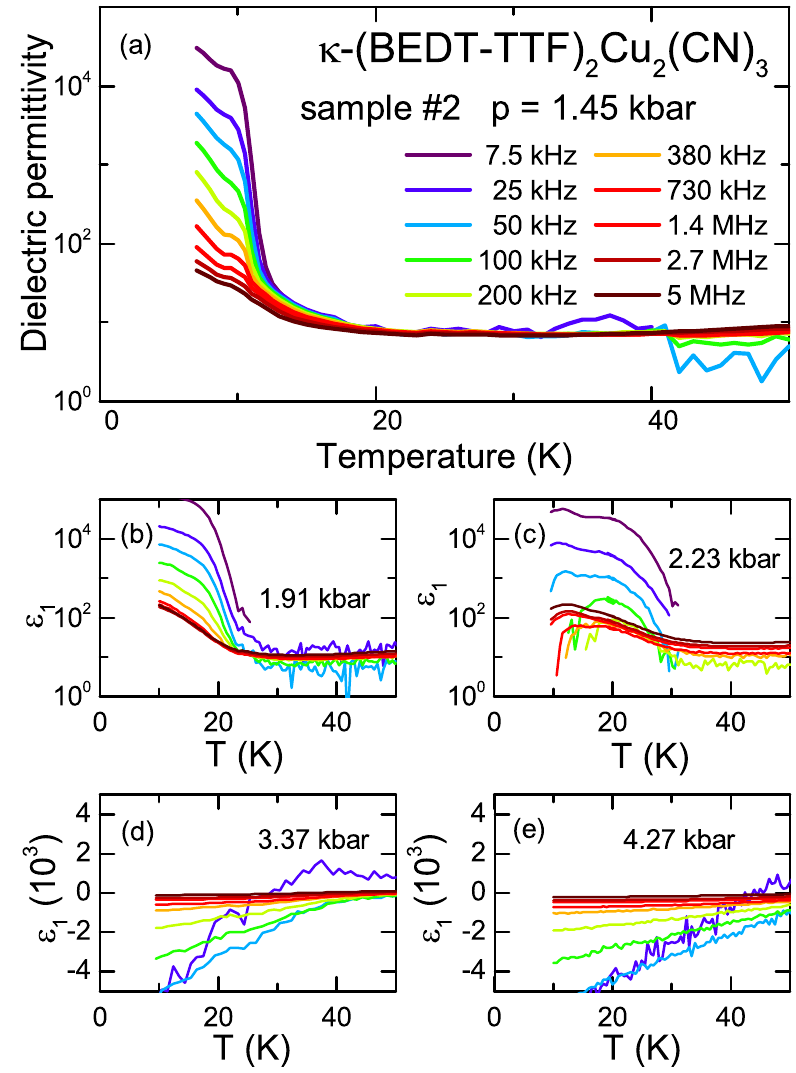}
	\caption{(a-c)~The dielectric permittivity of \CuCN\ is shown from $f=7.5$~kHz to 5~MHz in the coexistence regime between pressure values $p=1.45$ and 2.23~kbar. A strongly frequency-dependent, colossal enhancement of $\varepsilon_{1}(T)$ occurs at low temperatures, which is a result of spatial phase separation between metallic and insulating regions upon the first-order Mott IMT. (d,e)~$\varepsilon_{1}$ acquires negative values for $p\geq 3.37$~kbar indicating the onset of metallic transport. Note the different ordinates in panels (d,e). }
	\label{eps(T-high-p)}
\end{figure}

Besides these fundamental issues, there are some more peculiarities observed in \CuCN, which are far from being understood. Around $T\approx 6$~K the thermal expansion exhibits a pronounced anomaly \cite{Manna10,*Manna18} with related features observed in specific heat \cite{Yamashita08}, thermal conductivity \cite{Yamashita09}, ultrasound propagation \cite{Poirier14}, magnetic susceptibility \cite{Isono16}, and microwave dielectric properties \cite{Poirier12}.  Here, we add another striking phenomenology, presented in Fig.~\ref{Phase_diagram}, namely a collossal enhancement of the dielectric permittivity at the Mott IMT, reaching  up to $\varepsilon_{1}(T,p)\approx 10^5$ at the lowest frequencies (see Fig.~\ref{eps(T-high-p)}). This `dielectric catastrophe' is assigned to phase coexistence of spatially separated metallic and insulating regions at the first-order transition~\cite{Pustogow2021-percolation}. Note, this feature is distinct from the relaxor-type ferroelectric response (see Fig.~\ref{eps(T)}) that was observed at ambient pressure in the audio- and radio-frequency range below $T\approx 50$~K \cite{Abdel10,Pinteric14,*Pinteric15,*Pinteric18}. Previous attempts to link the latter to charge disproportionation within the dimers due to intersite Coulomb repulsion, dubbed quantum-electric dipole or paired-electron crystal \cite{Hotta10,*Hotta12,Naka10,Li10,*Dayal11,*Clay12,Gomi13}, could not be verified in experiment as
various spectroscopic methods have unanimously proven homogeneous charge distribution on the molecules \cite{Shimizu06,Sedlmeier12,Yakushi15,Tomic15,Pinteric16,Lasic18}.

%\section{Motivation}

In order to motivate our detailed experimental investigations and analysis, theoretical calculations and discussions, let us first give an overview on the tem\-per\-a\-ture-dependent dielectric response of \CuCN\ in the various regimes~\cite{Pustogow2021-percolation}.
In Fig.~\ref{eps(T)} we plot the real part of the dielectric permittivity $\varepsilon_{1}(T)$ for selected frequencies and pressures as indicated.
At $p=0$ and low pressures, the $T$ dependence of the permittivity is dominated by a pronounced peak. The maximum exhibits the characteristic frequency dependence of a relaxor ferroelectric and moves to lower temperatures with increasing pressure, as indicated by the dotted lines in Fig.~\ref{eps(T)}(a-e). Most prominent, however, are the dramatic changes of the dielectric response close to the IMT ($p_{\rm IMT}=1.45$~kbar) \cite{Furukawa18}. At low temperatures, $T<20$~K, the dielectric constant is strongly enhanced up to 2.2~kbar and acquires a frequency-dependent amplitude even exceeding $\varepsilon_{1}\approx 10^{5}$ at the lowest measured frequency [Fig.~\ref{eps(T-high-p)}(a-c)].
 We ascribe this observation to a phase coexistence around the Mott IMT, where metallic regions grow in an insulating matrix and eventually form percolating clusters through the sample.
For  $p\geq 3.37$~kbar [Fig.~\ref{eps(T-high-p)}(d,e)], $\varepsilon_{1}$ acquires large negative values as metallic behavior sets in.

After explaining the experiments, we analyze our observations on the insulating phase in detail, followed by the colossal permittivity enhancement in the transition region. In a next step, we present our theoretical modelling of the percolative Mott IMT, using a hybrid DMFT approach. Finally  our findings and new insight on the Mott transition are discussed comprehensively.

\section{Experimental details}
We synthesized high-quality \CuCN\ single crystals by standard electrochemical synthesis \cite{Geiser91,Komatsu96}; sample 1 was grown at the  Universit{\"a}t Stuttgart while sample 2 was prepared at Argonne National Laboratory. In this study we measured the complex electrical impedance as a function of pressure, temperature and frequency in order to obtain the complex conductivity $\hat{\sigma} = \sigma_{1} + {\rm i} \sigma_{2}$ or, equivalently, the permittivity $\hat{\varepsilon} = \varepsilon_{1} + {\rm i} \varepsilon_{2}$.
To that end, the crystals are contacted with thin gold wires that are attached by carbon paste to opposite surfaces of a single crystal, such that the measurements were performed out-of-plane with $E \perp bc$.
The experiments were carried out by measuring through two contacts in a pseudo four-point configuration \cite{Keysight} with an Agilent 4294 impedance analyzer.
To make sure that we operate in the Ohmic regime a small voltage of 0.5~V was applied.
In order to characterize the crystals, we have measured the low-frequency resistivity as a function of temperature and pressure.

Our pressure-dependent dielectric experiments were performed utilizing a piston-type pressure cell as described in detail in Refs.~\onlinecite{Roesslhuber18,Pustogow2021-percolation}. Using a self-made electrical feedthrough for coaxial cables, we could reach up to approximately 10~kbar. Daphne oil 7373 serves as the liquid pressure-transmitting medium, because it has a good hydrostaticity, is inert to molecular solids, and stays fluid at 300~K for pressures applied in this study.
An InSb semiconductor pressure gauge with negligible pressure gradient below $T=50$~K was employed for \textit{in-situ} determination of the inherent pressure loss upon cooling.
As a consequence, in the temperature range of particular interest here, the data are
collected in the same pressure cycles; this is important for comparison.
Unless indicated otherwise, throughout the manuscript we state the pressure reading at the lowest temperature $T=10$~K.

The pressure cell was cooled down in a custom-made continuous-flow helium cryostat
that allows us to reduce the total cable length to 50~cm enabling reliable measurements
at frequencies up to 5~MHz. The compact cryostat design results in a rather steep thermal gradient limiting the lowest reachable temperature to about 8~K. We kept the cooling rate below $0.4$~K/min for all measurements and observed no cooling-rate dependence. Since good agreement between the results of both samples was obtained, we present here the data of sample 2. The results obtained on sample 1 are added in Appendix \ref{sec:sample1}.

\section{Results and analysis}

\subsection{Dielectric response in the Mott insulating phase}
\label{sec:insulator}

At reduced temperatures the ambient-pressure dielectric constant $\varepsilon_{1}(T)$ of \CuCN\ reveals a peak, as first reported by Abdel-Jawad {\it et al.}
\cite{Abdel10} and later confirmed by Pinteri{\'c}  and collaborators \cite{Pinteric14,Pinteric15,Pinteric18}; here, we label this feature as high-temperature (HT) peak.
When probing with a frequency of $f=100$~kHz, for instance,
the maximum appears at $T=28$~K, in other crystals up to 40~K, in agreement with previous reports. The observed sample dependence \cite{Pinteric14} is confirmed by disorder studies, which reveal a shift of the maximum in $\varepsilon_{1}(T)$ to lower temperatures upon x-ray irradiation \cite{Sasaki15}.
Fig.~\ref{eps(T)}(a) illustrates how the maximum
moves to low temperatures when probed at smaller frequencies;
at the same time, however, it gets more pronounced.
This behavior resembles the well-known phenomenology of relaxor ferroelectrics \cite{Cross08}.

A closer look reveals a shoulder-like feature around $T=15$~K, which we denote as low-temperature (LT) mode;
it evolves into a small second peak for frequencies between $f=53$ and 200~kHz. As pressure rises, this LT mode becomes a well-defined peak, it grows in amplitude and eventually dominates the spectrum at $p=0.86$~kbar, as seen in Fig.~\ref{eps(T)}(c).
Although the HT feature seems to maintain its amplitude and width, it becomes secondary.
%Interestingly, this LT peak is visible in sample \#2 already at ambient pressure whereas its presence becomes obvious in sample \#1 upon pressure.
%This might also be the case in previous ambient pressure studies \cite{Abdel2010,Pinteric2014,Sasaki2015,Pinteric2018} where a corresponding feature is not observed.
Both modes shift to lower temperatures with pressure (cf.\ Fig.~\ref{T_B}).
%The frequency-dependent shift to higher temperatures persists up to 4.65~kbar for the HT peak, whereas the LT peak seems to be non-dispersive. Here we want to point out, that a detailed investigation of the dispersive behavior in dependence of pressure by determining the peak position for all frequencies is difficult since both peaks overlap for the majority of the investigated frequencies.

\begin{figure}[b]
	\centering
	\includegraphics[width=0.7\columnwidth]{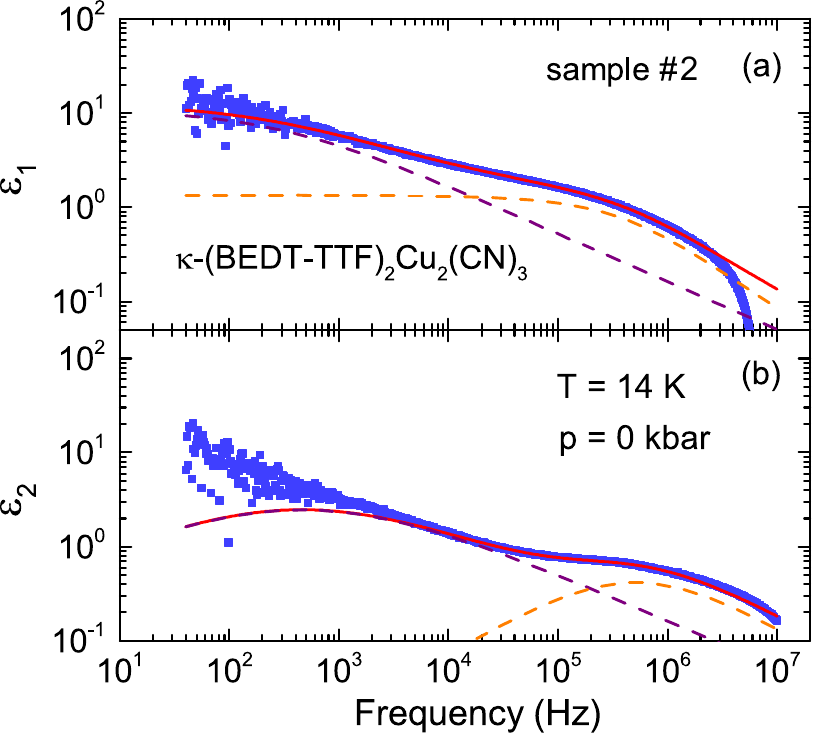}
	\caption{The real and imaginary parts of the dielectric constant of \CuCN, $\varepsilon_{1}(f)$ and $\varepsilon_{2}(f)$, as a function of frequency for $T=14$~K at ambient pressure. The dashed lines represent two Cole-Cole modes; the full lines correspond to their sum of both, according to Eq.~(\ref{Cole-Cole}).}
	\label{eps(f)}
\end{figure}

For a better understanding of the physical background, we analyze the
frequency-dependence of the dielectric response, plotted in Fig.~\ref{eps(f)} for the example of $T=14$~K and $p=0$~kbar.
We should note that even at these low temperatures the resistivity of $\kappa$-(BEDT-TTF)$_2$Cu$_2$(CN)$_3$ remains at moderate values;
implying that the dc-conductivity $\sigma_{dc}$ gives a considerable contribution to the imaginary part of the permittivity. Following the common procedure, we subtract this part: $\varepsilon_{2}(\omega) =  \left[\sigma_{1}(\omega) - \sigma_{dc}\right]/\omega\varepsilon_{0}$ \cite{Staresinic2006a,Pinteric14}. In Fig.~\ref{eps(f)} we can distinguish two relaxation modes as roll-offs in the real part $\varepsilon_{1}(f)$ and broad maxima in the imaginary part $\varepsilon_{2}(f)$.
Hence we fit our data by the sum of two Cole-Cole modes:
\begin{equation}
\hat{\varepsilon}(\omega) -\varepsilon_{\infty} = \frac{\Delta \varepsilon^{\text{mode 1}}}{1 + ({\rm i} \omega \tau_{1})^{1-\alpha_{1}}} + \frac{\Delta \varepsilon^{\text{mode 2}}}{1 + ({\rm i} \omega \tau_{2})^{1-\alpha_{2}}} \quad , \label{Cole-Cole}
\end{equation}
wherein $\tau_{1,2}$ are the relaxation times, $\omega=2\pi f$ the angular frequency of the applied electric ac-field, $(1-\alpha_{1,2})$ are real-valued and the parameters describing the symmetric broadening of the relaxation time distribution functions, $\Delta \varepsilon^{\text{mode 1}}$ and  $\Delta \varepsilon^{\text{mode 2}}$ are real-valued and denote the dielectric strengths of the corresponding modes, with $\Delta \varepsilon^{\text{mode 1}} + \Delta \varepsilon^{\text{mode 2}} = \varepsilon_{\rm static}  - \varepsilon_{\infty}$, wherein $\varepsilon_{\rm static}$ and $\varepsilon_{\infty}$ are the real values for low and high frequencies, respectively.
Ambient-pressure studies \cite{Pinteric14} previously identified only a single mode; similar to the related compounds $\kappa$-(BEDT-TTF)$_2$Ag$_2$(CN)$_3$ \cite{Pinteric16}.

\begin{figure}
	\centering \includegraphics[width=0.7\columnwidth]{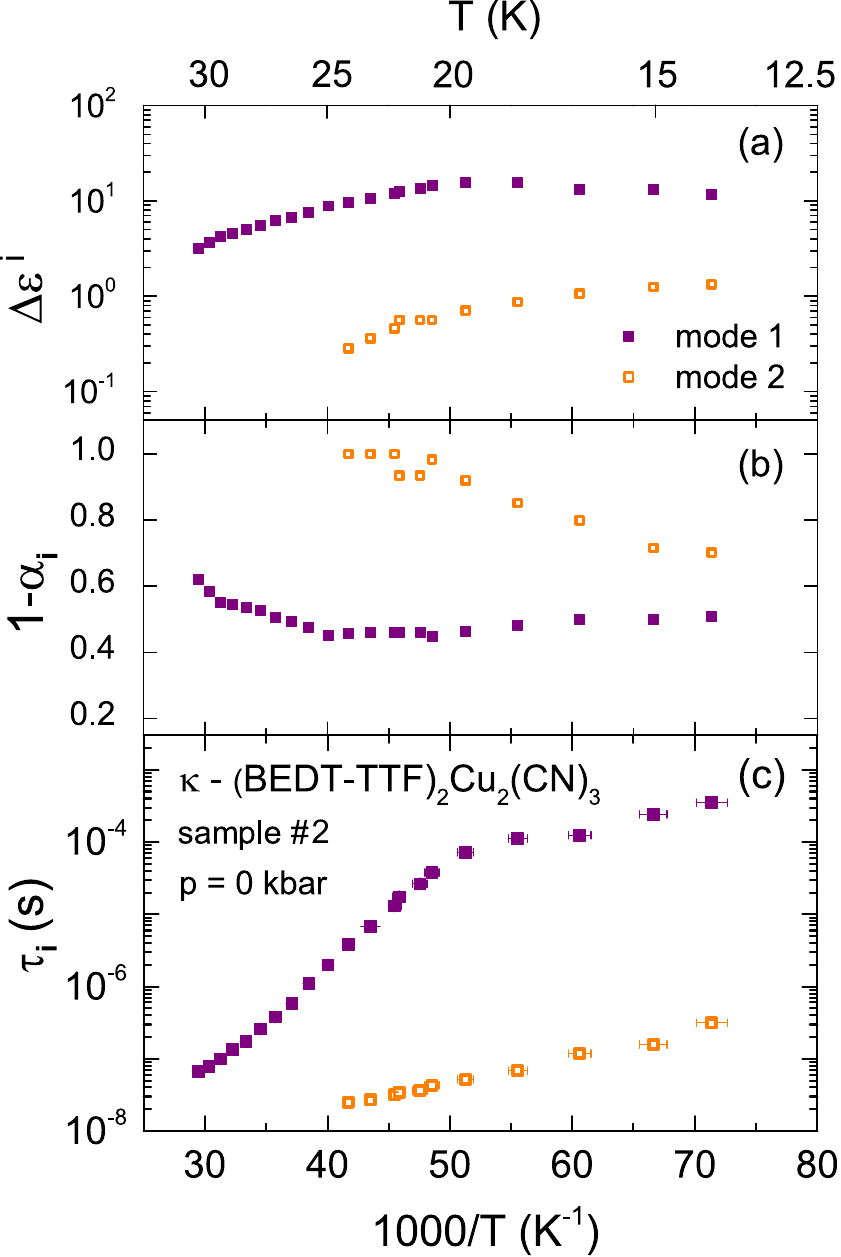}
	\caption{Arrhenius plots of (a)~the dielectric strength $\Delta \varepsilon^{\text{mode 1,2}}(T)$, (b)~the distribution of relaxation times $1-\alpha(T)$ and (c)~the mean relaxation time $\tau(T)$ for both modes obtained from the fits of dielectric data of \CuCN\ measured at  ambient pressure. The solid purple symbols refer to mode 1 while the open orange squares indicate the data of mode 2.}
	\label{Debye_1_2_0kbar}
\end{figure}

The parameters obtained from fitting the ambient-pressure data are plotted in Fig.~\ref{Debye_1_2_0kbar} as a function of inverse temperature.
The strength of the first mode $\Delta \varepsilon^{\text{mode 1}}(T)$ shows a peak around $T=22$~K, resembling $\varepsilon_{1}(T)$ for low frequencies. The second contribution $\Delta \varepsilon^{\text{mode 2}}$ is approximately one order of magnitude smaller, but increases monotonously upon cooling. With reducing the temperature, $1-\alpha_{1}(T)$ decreases and the relaxation time $\tau_{1}(T)$ increases, providing evidence
for significant broadening and slowing down of the dielectric relaxation; in relaxor ferroelectrics this is usually ascribed to cooperative motion and glassy freezing \cite{Cross08}. Interestingly, we observe a kink in $\tau_{1}$ and a concomitant slight increase of $1-\alpha_{1}$ around $T=20$~K. For the second mode, $\tau_{2}$ becomes continuously  larger as the temperature is reduced, and $1-\alpha_{2}$ decreases; again indicating that the corresponding relaxation freezes out.

\begin{figure}[b]
	\centering
	\includegraphics[width=0.7\columnwidth]{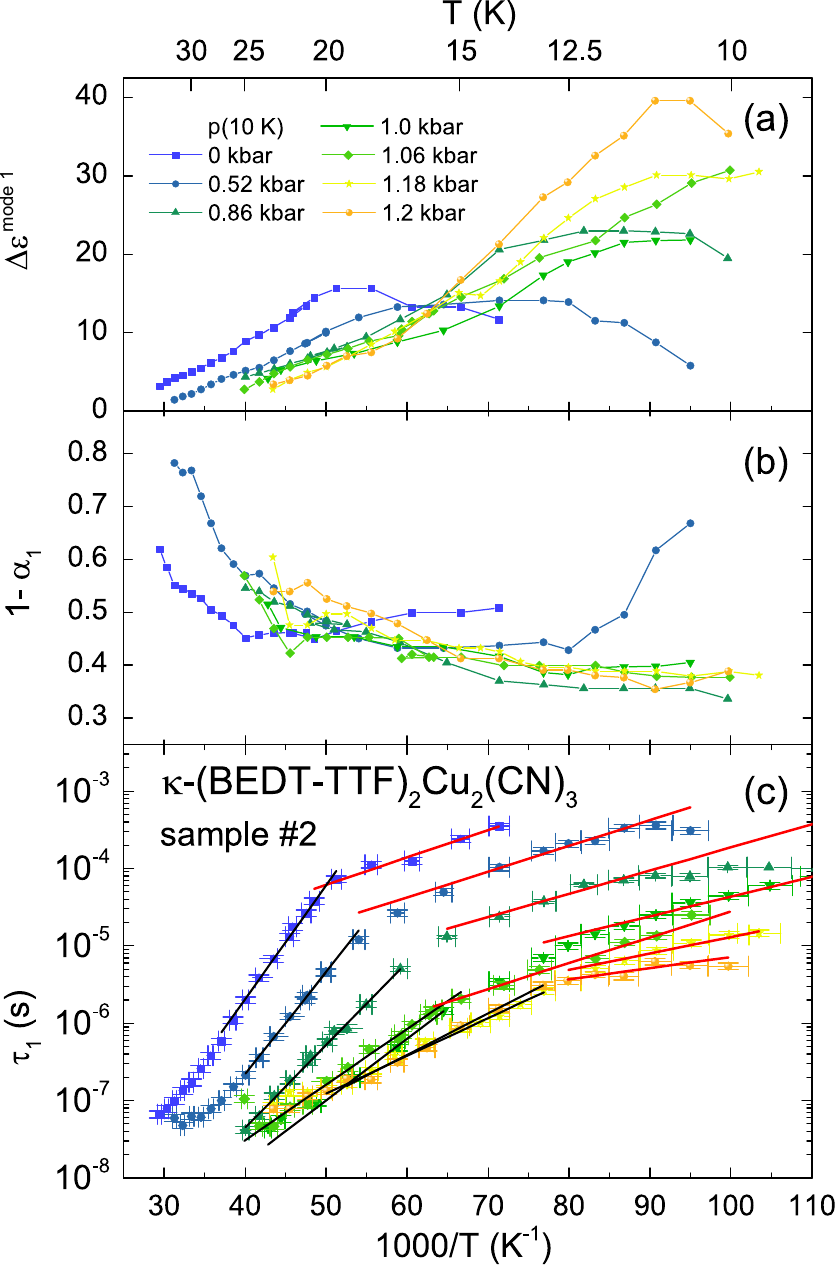}
	\caption{Arrhenius plots of the Debye parameters of the low-frequency mode 1 in \CuCN\ for pressures up to 1.22~kbar. (a)~Dielectric strength $\Delta \varepsilon^{\text{mode 1}}(T)$, (b)~distribution of relaxation times $1-\alpha_{1}(T)$ and (c)~mean relaxation time $\tau_{1}(T)$. The black and red lines represent fits with Eq.~(\ref{eq:HT}) above and below the kink in $\tau_{1}(T)$ at $T_{\rm B}$, respectively.}
	\label{Debye_1}
\end{figure}

\subsubsection{Analysis of Low-Frequency Mode 1}
\label{sec:Mode1}

When turning to the pressure dependence, Fig.~\ref{Debye_1} displays the fit parameters corresponding to mode~1 as a function of inverse temperature; the presentation is limited to the insulating state with $p\leq 1.2$~kbar.
The peak in dielectric strength $\Delta \varepsilon^{\text{mode 1}}(T)$ shifts to lower temperatures and increases in amplitude as pressure is applied. This corresponds to the evolution of $\varepsilon_{1}(T)$ plotted in Fig.~\ref{eps(T)},
where the appearance of the LT feature broadens the relaxation.
The resulting lower values of $1-\alpha_{1}$ do not indicate
more cooperativity or glassy behavior compared to ambient pressure.
The kink in $\tau_{1}(T)$ broadens with pressure and shifts to lower temperatures, the corresponding relaxation time gets shorter.
A similar feature was previously observed at ambient pressure around $T=17$~K
and attributed to a bifurcation temperature $T_{\rm B}$ \cite{Pinteric14}; here free charge carriers start to freeze out and hopping-like conduction sets in.

\begin{figure}
	\centering
	\includegraphics[width=0.87\columnwidth]{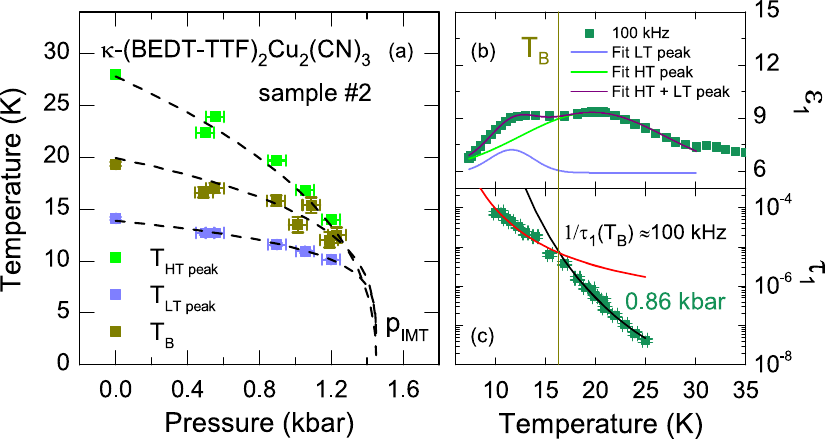}
	\caption{(a)~Pressure dependence of the bifurcation temperature $T_{\rm B}$ and the positions of the high and low-temperature peak, $T_{\rm HT}$ and $T_{\rm LT}$, in \CuCN; here $p_{\rm IMT}$=1.45~kbar denotes the critical pressure of the insulator-to-metal transition \cite{Geiser91,Furukawa18}.
		(b)~For the example of $p=0.86$~kbar and $f=100$~kHz the temperature dependence of the dielectric constant $\varepsilon_{1}(T)$ is plotted together with the fits of the HT and LT peak by two Gaussian functions (blue and green lines). (c) The relaxation time $\tau_{1}(T)$ with the fits from Fig.~\ref{Debye_1}(c) according to Eq.~(\ref{eq:HT}) for the temperature range above (black line) and below (red line) the kink at $T_{\rm B}$. The crossover from the HT to the LT peak is located at $T_{\rm B}$.
		%with a concomitant change in relaxation dynamics.
	}
	\label{T_B}
\end{figure}

%Let us examine $T_{\rm B}$ before we discuss the pressure-dependence of $\tau_{1}(T)$ in detail.
In Fig.~\ref{T_B}(b,c), $\varepsilon_{1}(T)$ and $\tau_{1}(T)$ for $f=100$~kHz are plotted as a function of temperature at $p=0.86$~kbar. The HT and LT features in $\varepsilon_{1}(T)$ are well described by two Gaussian functions.
Now it becomes clear that the kink in $\tau_{1}(T)$ at $T_{\rm B}$ (indicated by solid green circles in Fig.~\ref{Phase_diagram}) corresponds to the transition from the HT peak to the LT peak with a concomitant change in the relaxation mechanism. We apply this procedure to all pressures and plot the pressure evolution of $T_{\rm B}$ in panel (a) of Fig.~\ref{T_B} together with the positions of the peaks at $f=100$~kHz. The dashed lines represent extrapolations to $p_{\rm IMT} = 1.45$~kbar \cite{Geiser91,Furukawa18} according to
\begin{equation}
\frac{T_{i}(p)}{T_{i}^0} =
 \left( \frac{p_{\rm IMT} - p}{p_{\rm IMT}} \right)^{z} \quad, \label{eq:IMT}
\end{equation}
with $i = \left\{\mathrm{HT,B,LT} \right\}$ the three characteristic temperatures,
$T_{i}^0$ giving a intercept at $p=0$ and $z$ the critical exponent. The obtained parameters are listed in Table~\ref{tab:IMT}.

\begin{table}[t]
	\caption{Parameters of \CuCN~obtained from fits of the pressure evolution of the high-temperature peak $T_{\rm HT}$, the bifurcation temperatures $T_{\rm B}$ and the low-temperature peak $T_{\rm LT}$ according to Eq.~(\ref{eq:IMT}) with $p_{\rm IMT}$=1.45~kbar.
$T_i^0$ gives the respective ambient-pressure temperature and $z$ is a unitless exponent. }
	\centering
	\begin{tabular}{c|ccc}
		 &  $T_{\rm LT}$  & $T_{\rm B}$ & $T_{\rm HT}$   \\
		\hline
		$T_{i}^0$ (K)& 13.0$\pm$ 0.1 &  19.9 $\pm$ 0.2 & 27.7 $\pm$ 0.5  \\
		$z$ &  0.18 $\pm$ 0.01 & 0.26 $\pm$ 0.02& 0.41 $\pm$ 0.08
		\label{tab:IMT}
	\end{tabular}
\end{table}

For $T>T_{\rm B}$, we can describe the dependence of the relaxation time $\tau_{1}$ on temperature by an activated behavior
\begin{equation}
\tau_{1} = \tau_{\rm HT} \exp\left\{ {\Delta_{\rm HT} }/{T} \right\}  \quad ,
  \label{eq:HT}
\end{equation}
which is represented by the black  line in Fig.~\ref{Debye_1}(c).
%In general, a decrease of the activation energy indicates that the entities responsible for the relaxation are easier to flip, deform or shift and the high-temperature limit time scale can be considered as a measure for the size of these entities.
Also the low-temperature regime can be fitted with an activated behavior in analogy to Eq.~(\ref{eq:HT}), which is illustrated by the red line in Fig.~\ref{Debye_1}(c).

\begin{figure}[b]
	\centering
	\includegraphics[width=0.9\columnwidth]{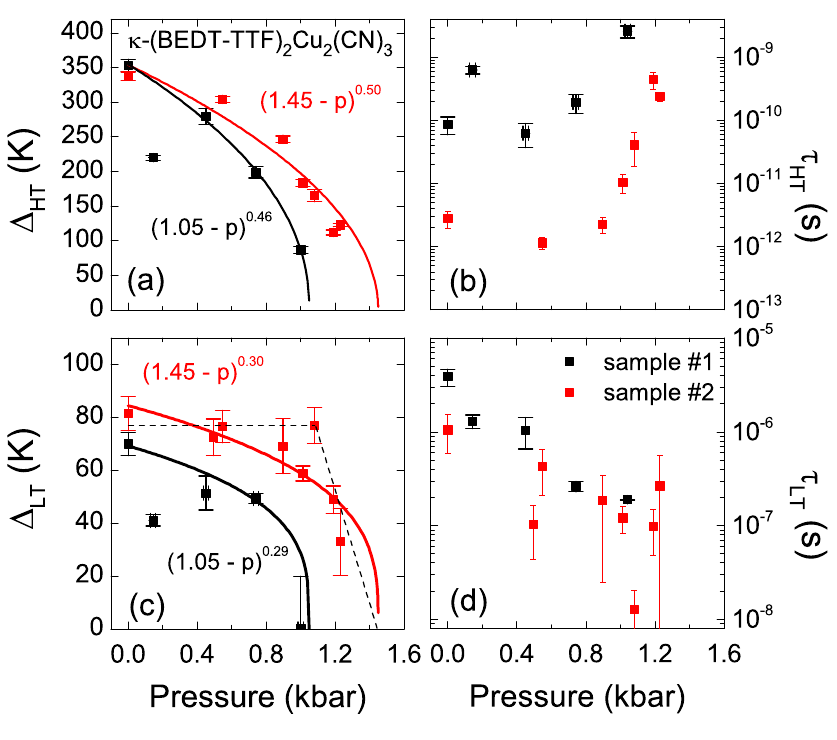}
	\caption{Pressure dependence of the activation energy $\Delta$ and relaxation time $\tau$ of \CuCN ~obtained from fits of the corresponding temperature behavior of $\tau_{1}(T)$ shown in Fig.~\ref{Debye_1}(c). The upper panels (a,b) correspond to the high-temperature regime; the lower panels (c,d) to the low-temperature regime. The black symbols correspond to sample~1, the red symbols to sample~2 while the solid lines in (a,c) represent mean-field fits in analogy to Eq.~\ref{eq:IMT}.  Most importantly, $\Delta_{\rm LT}$ is rather constant and abruptly decreases around 1.1~kbar, reminiscent of a first-order phase transition with $p_{\rm IMT} = 1.45$~kbar.}
	\label{Fit_tau}
\end{figure}

The extracted fit parameters for the HT mode ($T>T_{\rm B}$) and the LT mode ($T<T_{\rm B}$) are plotted in Fig.~\ref{Fit_tau} as a function of pressure; here we also include the qualitatively similar results obtained from sample 1 while the corresponding plots are presented in the Appendix \ref{sec:sample1}. Our findings are in line with previous ambient-pressure studies \cite{Pinteric14}. $\Delta_{\rm HT}(p)$ decreases as pressure rises following a mean-field behavior in analogy to Eq.~\ref{eq:IMT}. Best fits were obtained by using fixed $p_{\rm{IMT,2}}=1.45$~kbar and $p_{\rm{IMT,1}}=1.05$~kbar for sample 2 and sample 1, respectively. The fit parameters are summarized in Table~\ref{tab:Fit_Delta} and show good agreement between the two samples for the critical exponents.

\begin{table}
	\caption{Mean field parameters $\Delta_{i}^0$ and $z$ of $\Delta_{i}(p)$ in analogy to Eq.~(\ref{eq:IMT}) using fixed $p_{\rm{IMT,2}}$=1.45~kbar for sample~2 and $p_{\rm{IMT,1}}=1.05$~kbar for sample~1.}
	\centering
	\begin{tabular}{c|cc}
		sample 2&  $\Delta_{\rm{HT}}$ & $\Delta_{\rm{LT}}$    \\
		\hline
		 $\Delta_{i}^0$  (K) & 354  $\pm$  19   &  84 $\pm$ 2   \\
		$z$                  & 0.50 $\pm$ 0.09  & 0.30 $\pm$ 0.04\\
%	\end{tabular}
\multicolumn{3}{c}{ }\\
%	\quad
%	\begin{tabular}{c|cc}
		sample 1 &  $\Delta_{\rm{HT}}$ & $\Delta_{\rm{LT}}$    \\
		\hline
		$\Delta_{i}^0$  (K)  &  355 $\pm$ 5 & 69 $\pm$ 4  \\
		$z$ & 0.46 $\pm$ 0.03  & 0.29 $\pm$ 0.05
		%\bottomrule
	\end{tabular}	
	\label{tab:Fit_Delta}
\end{table}

$\Delta_{\rm LT}(p)$ [Fig.~\ref{Fit_tau}(c)] is basically pressure-independent up to $1.1$~kbar, followed by a strong decrease when $p$ is further increased. The behavior is reminiscent of a first-order phase transition and extrapolates to $\Delta_{\rm LT}=0$ at $p_{\rm IMT}=1.45$~kbar; in excellent agreement with the findings in Fig.~\ref{T_B}. For sample 1, it is difficult to pin down the pressure dependence of $\Delta_{\rm{LT}}$ because the number of data points is limited. Since $\Delta_{\rm{HT}}$ and $T_{\rm{B}}$ both follow mean-field behavior, we also apply rough tests for mean-field behavior on $\Delta_{\rm LT}$, represented by the solid lines.

Apart from a rather strong sample dependence, the time scale $\tau_{\rm HT}$ [Fig.~\ref{Fit_tau}(b)] becomes longer upon applying pressure. On the other hand, $\tau_{\rm LT}$ [Fig.~\ref{Fit_tau}(d)] decreases upon applying pressure and sample dependence is less pronounced.

\subsubsection{Analysis of High-Frequency Mode 2}
\begin{figure}
	\centering
	\includegraphics[width=0.7\columnwidth]{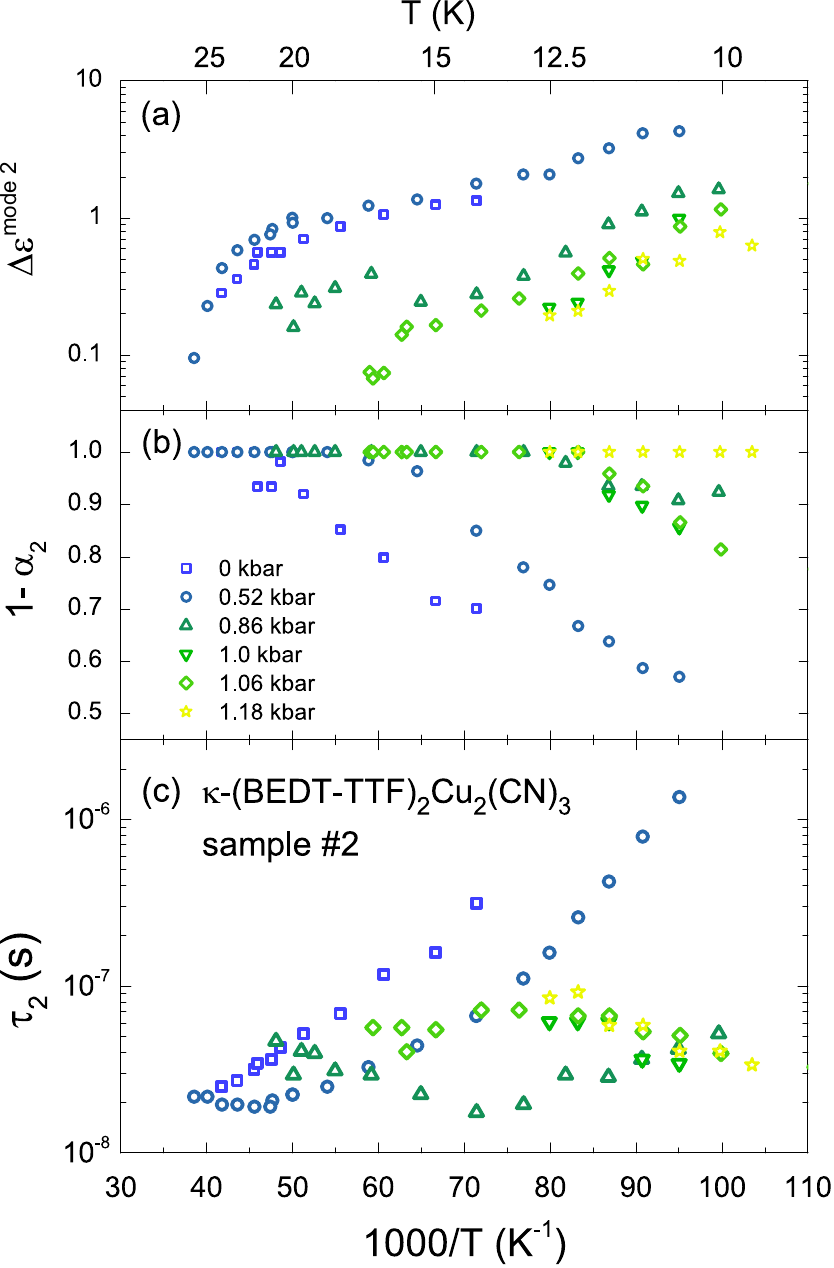}
	\caption{Temperature dependence of the parameters describing mode 2 in \CuCN\ at various pressures. (a) Dielectric strength $\Delta \varepsilon^{\text{mode 2}}$, (b) distribution of relaxation times $1-\alpha_{2}$ and (c) mean relaxation time $\tau_{2}$ in an Arrhenius plot versus inverse temperature. Upon increasing pressure, a pronounced minimum in $\tau_{2}$ develops, indicating a non-monotonic relaxation dynamics in \CuCN.}
	\label{Debye2}
\end{figure}

In the Arrhenius plot of Fig.~\ref{Debye2} we show the parameters of the second mode: $\Delta \varepsilon_{2}(T)$, $1-\alpha_{2}(T)$ and $\tau_{2}(T)$, as obtained from the Cole-Cole fits. Throughout the whole pressure-temperature range wherein mode~2 is observed,
its strength $\Delta \varepsilon^{\text{mode 2}}$ stays an order of magnitude below the one of the low-frequency mode~1.
$\Delta \varepsilon^{\text{mode 2}}(T)$ increases steadily as the temperature is reduced.
With rising pressure the mode shifts to lower temperatures (Fig.~\ref{eps(T)}) and correspondingly does the enhancement in
$\Delta \varepsilon^{\text{mode 2}}(T)$.
A drop in $1-\alpha_{2}(T)$ indicates a considerable broadening of the mode. The temperature dependence of $\tau_{2}$ is strongly influenced by pressure.
While at ambient pressure $\tau_{2}(T)$ monotonically increases,
at $p=0.52$~kbar a minimum is observed that shifts to lower temperatures and gets more pronounced at $p=0.86$~kbar.
In other words, $\tau_{2}(T)$ depends on pressure in a non-monotonic way.
A similar relaxation dynamics has been widely observed for confined systems as well as in the relaxor ferroelectric KTa$_{0.65}$Nb$_{0.35}$O$_{3}$ when doped with Cu by approximately 0.1\%\ \cite{Ryabov04}.

\subsection{Dielectric response at the insulator-metal transition}
\label{sec:percolation}

While the relaxor-ferroelectric response from Fig.~\ref{eps(T)} has been subject to much controversy and debate\cite{Abdel10,Pinteric14}, the most striking observation of our present study is the colossal enhancement of the dielectric constant around the first-order Mott IMT, displayed in Figs.~\ref{Phase_diagram} and \ref{eps(T-high-p)}(a-c). At low temperatures, $\varepsilon_1$ acquires values of several hundred up to $10^5$ -- increasing towards low frequencies -- in the pressure region 1.45--2.23~kbar, with a peak around 1.8~kbar at $T=10$~K. Our thorough analysis of the dielectric relaxation in Sec.~\ref{sec:insulator} clearly shows that this behavior has a distinct origin. We interpret this phenomenology as a result of percolating metallic clusters embedded in an insulating matrix~\cite{Pustogow2021-percolation} -- a situation similar to other systems subject to a metal-insulator transition~\cite{Qazilbash07,Hovel10}. On these grounds, we now estimate the pressure dependence of the filling fraction from our experimental data.

%\subsubsection{Estimate of the metallic filling fraction from experimental results}

\begin{figure}
	\centering
	\includegraphics[width=0.7\columnwidth]{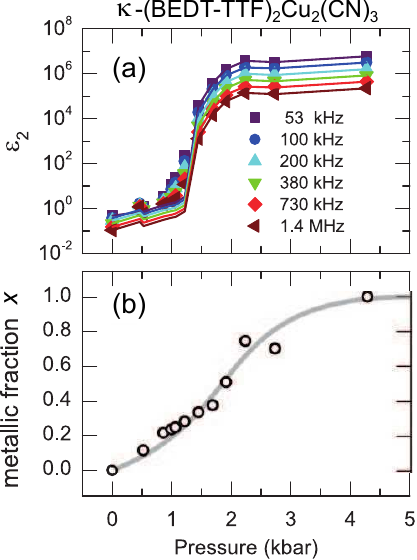}
	\caption{(a) Pressure dependence of $\varepsilon_2$ of \CuCN\ measured at $T=10$~K using different frequencies as indicated. The solid lines represent fits by Bruggeman's effective medium approximation [Eq.~(\ref{eq:BEMA})] for spherical inclusions with the metallic filling fraction as free parameter. (b) Consistent for all frequencies, the metallic fraction $x$ in dependence of pressure exhibits a rapid change around percolation; the grey line is a guide to the eye.}
	\label{BEMA}
\end{figure}

If the pressure-driven IMT in our Mott system is a first-order transition for $T < T_{\rm crit}$,
two phases are thermodynamically stable in the coexistence region \cite{Georges96,Vucicevic13}:
One represents the metallic state while the other corresponds to the insulating phase.
In the presence of weak disorder, such a region will feature a mixture of randomly distributed metallic and insulating domains,
with respective volume fractions that vary with pressure.
The IMT takes place when the volume fraction of the metallic phase approaches the percolation threshold.
The dielectric properties of  such a mixture can be modeled by
Bruggeman's effective medium approximation (BEMA) \cite{Stroud75,Stroud78,Kirkpatrick73,Choy15}:
\begin{equation}
x \, \frac{\varepsilon_{m}-\varepsilon_{\rm eff}}{\varepsilon_{\rm eff} + L (\varepsilon_{m}-\varepsilon_{\rm eff})} + (1-x) \, \frac{\varepsilon_{i}-\varepsilon_{\rm eff}}{\varepsilon_{\rm eff} + L (\varepsilon_{i}-\varepsilon_{\rm eff})} = 0  \quad ,\label{eq:BEMA}
\end{equation}
where $x$ is the volume fraction of the metallic inclusions,
$L$ is the shape factor,
$\varepsilon_{i}$ and $\varepsilon_{\rm m}$ are the complex permittivities of the insulating and metallic phases, respectively, and $\varepsilon_{\rm eff}$ is the effective permittivity of the composite.
In Fig.~\ref{BEMA}(a) the imaginary part of the dielectric constant of \CuCN\ is plotted as a function of pressure measured at various frequencies at the lowest accessible temperature. We find the step in $\varepsilon_2(p)=\sigma_1(p)/(\varepsilon_0 \omega)$ becoming less pronounced with increasing frequencies.

In order to estimate the metallic filling fraction $x$ from these data, we assume that for $p=0$ the specimen is completely in the insulating phase, the properties resemble
$\varepsilon_i$; for $p=4.3$~kbar the metallic state is fully established, corresponding to $\varepsilon_m$.
For each particular pressure, we can now obtain a value of  $x$ that best describes the experimental data by  the effective permittivity $\varepsilon_{\rm eff}$ calculated via Eq.~(\ref{eq:BEMA}) assuming spherical inclusions: $L=1/3$. In Fig.~\ref{BEMA}(a) the respective fits are shown by solid lines for the various frequencies.
The resulting filling fraction $x$ is plotted in Fig.~\ref{BEMA}(b) as a function of pressure.
The gradual increase follows a $\tanh$-like behavior around the IMT, supporting our assumption for the theoretical simulations \cite{Pustogow2021-percolation}.

We expect this crude method to severely overestimate the extent of the coexistence regime because the pressure dependence of the conduction properties is neglected. A narrower region of around 1 to 2~kbar is more likely. In Section~\ref{theory-background} we address this issue by a phenomenological model utilizing an approach of hybrid dynamical mean-field theory.

\begin{figure}
	\centering
	\includegraphics[width=1.0\columnwidth]{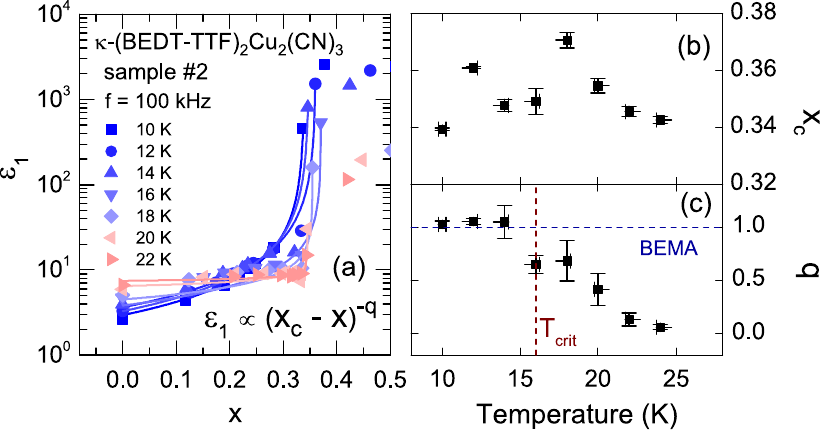}
	\caption{(a)~Dielectric constant $\varepsilon_{1}$ of \CuCN~ measured at $f=100$~kHz for several temperatures in dependence of the relative pressure $(p-p_c)$. The solid lines represent fits according to Eq.~\ref{eq:eps_q}.~~(b)~Temperature dependence of the percolation threshold $p_c$, as obtained from the fits in (a).~~(c)~Temperature dependence of the exponents $q$. We attribute the drop of $q$ above $T_{\rm crit}$ to the change from the first-order insulator-metal transition to a crossover at higher temperatures. For clarity reasons, we also include $T_{\rm crit}=16$~K (dashed red line) and the predictions for $q$ according to the BEMA model (dashed blue line).}
	\label{Fit_eps}
\end{figure}
Now that we obtained the pressure-dependence of the metallic filling fraction $x(p)$,
we are in the position to analyze $\varepsilon_{1}(x)$. The corresponding plot for the 100~kHz data
is shown in Fig.~\ref{Fit_eps}(a).
The static dielectric constant of a percolating system is a function of $x$ and its divergence at the percolation threshold $x_{c}$ can be described by~\cite{Efros76}
\begin{equation}
\varepsilon_{1}(\omega \rightarrow 0,T \rightarrow 0,x) \propto (x_{c} - x)^{-q}  \quad , \label{eq:eps_q}
\end{equation}
where the critical exponent $q$ depends on the dimension of the system; in three dimensions we expect $q$ ranging from 0.8 to 1 \cite{Efros76,Benguigui85,Clerc90,Hovel10,*Hovel11}, while $q=1.3$ is calculated for two dimensions \cite{Efros76}. The BEMA model predicts $q = 1$ independent of the dimensionality of the percolating system \cite{Clerc90}.

The temperature dependence of the percolation threshold $x_c$ and the exponent $q$ obtained from the fits (solid lines) in Fig.~\ref{Fit_eps}(a) is presented in panels (b) and (c), respectively. Percolation is established at a critical filling fraction $x_c\approx 0.35$, in excellent agreement with the prediction of ${1}/{3}$ for a three-dimensional system.
From our pressure-dependent dielectric measurements we find $q \approx 1$ up to $T_{\rm crit}$ as expected within the BEMA framework.

\subsection{Theoretical analysis of the dielectric permittivity at the IMT}
\label{theory-background}

\begin{figure*}
	\centering
	\includegraphics[width=1\textwidth]{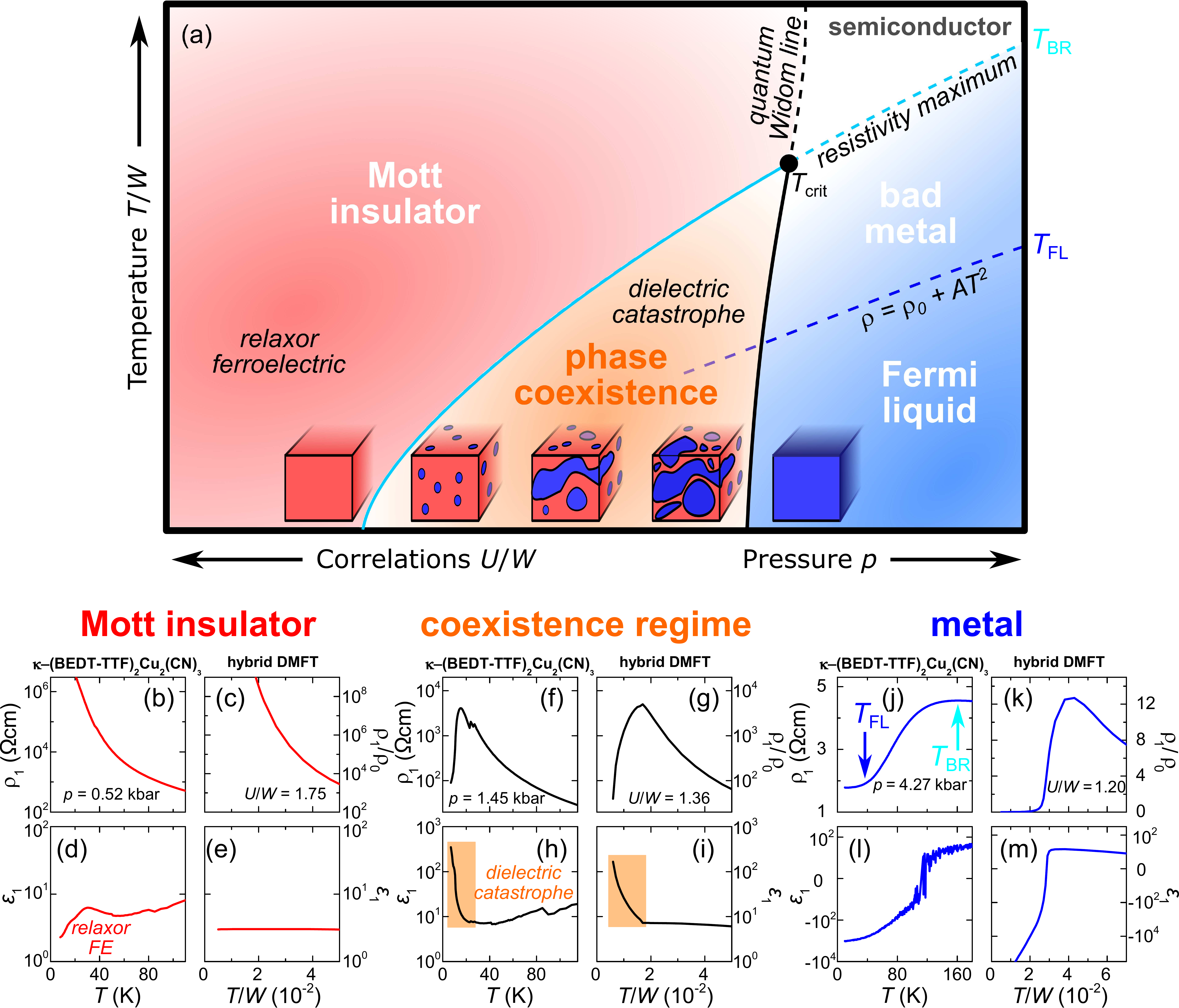}
	\caption{(a) Phase diagram of $genuine$ Mott insulators with first-order insulator-metal coexistence indicated at low temperatures. The bottom panels are grouped to illustrate the responses of the respective regimes, comparing our  experiments on \CuCN\ and our hybrid DMFT calculations. (b-e) The Mott-insulating state yields thermally activated resistivity and small, postive values of the permittivity. (f-g) While $\rho_1$ indicates a reduction with cooling when metallic clusters percolate, $\varepsilon_{1}$ is strongly increased upon metal-insulator phase coexistence. (j-m) The correlated metallic state below the Brinkman-Rice temperature $T_{\rm BR}$ exhibits Fermi-liquid properties with a quadratic temperature dependence of the resistivity at low temperatures, accompanied by large negative values of $\varepsilon_{1}$. The dielectric permittivity $\varepsilon_1$ was measured at $f=380$~kHz and calculated for $hf/W=5\times 10^{-9}$. Note, there is no residual resistivity in the theory data plotted in panel (k).
	}
	\label{theory}
\end{figure*}

We have just established a way to estimate $\varepsilon_{\rm eff}$ of this percolative first-order insulator-metal transition (IMT)
provided we know the properties of the constituting phases.
Now we want go this avenue on purely theoretical ground. In particular, we determine the complex permittivity of the mixed phase by calculating first the pure phases, together with their full correlation dependence, and then embedding them in a percolating network according to Eq.~(\ref{BEMA}). The technical details and the results of our single-site dynamical mean-field theory (DMFT) calculations of the optical conductivity~\cite{Georges96,Economou06,Letfulov01} are described in full length in Ref.~\onlinecite{Pustogow2021-percolation}. The resulting resistivities and permittivities  in the dielectric range ($hf/W=5\cdot 10^{-9}$) are compared to exemplary experimental data in Fig.~\ref{theory}. Panel (a) summarizes the three main regimes of the phase diagram, with the corresponding $\rho_1(T)$ and $\varepsilon_1(T)$ curves from experiment (left) and hybrid DMFT (right) placed below each other.

Fig.~\ref{theory}(b-e) illustrate the dielectric response in the homogeneous (i.e. $x=0$) Mott-insulating state. While the resistivity exhibits a monotonic increase upon cooling, the permittivity acquires small values between 1--10 typical of a charge-gapped state. In the coexistence regime around the IMT, plotted in panels (f-i), the temperature dependence  of $\rho(T)$ becomes metal-like as the filling fraction exceeds the percolation threshold, forming a conducting path through the entire sample [see bottom sketches of frame (a)]. This is accompanied by a collossal enhancement of $\varepsilon_1$ by several orders of magnitude at low temperatures; the `dielectric catastrophe' is well reproduced by our hybrid DMFT calculations. Also the response of the pure metallic state ($x=1$) in Fig.~\ref{theory}(j-m) is captured well by theory. Here, Fermi-liquid behavior with $\rho_1\propto AT^2$ below $T_{\rm FL}$ is followed by a bad metallic state extending up to the Brinkman-Rice temperature $T_{\rm BR}$ where quasiparticles are ultimately destroyed~\cite{Radonjic2012,Deng2013,Pustogow2021-Landau}. Accordingly, the dielectric constant turns negative for $T<T_{\rm BR}$ as expected for a metal.

Remarkably, our hybrid DMFT approach succeeds in modeling the main features of low-frequency complex transport properties throughout the Mott IMT, supporting our conclusions based upon optical conductivity work that the main physics of \CuCN\ is captured by the single-band Hubbard model~\cite{Pustogow18b}.
Having said that, we note that the measured relaxor-ferroelectric behavior in the Mott-insulating state is not reproduced by theory, indicating that it does not originate from intrinsic Mott physics. This is corroborated by a pronounced sample dependence indicative of the relevance of impurities~\cite{Pinteric14}.

\section{Discussion}
\label{discussion}
\subsection{Dielectric relaxation in the insulating state}
\subsubsection{High-temperature peak}

Our pressure-dependent investigations unveil that the dielectric response of \CuCN ~contains two dielectric contributions. The high-temperature (HT) peak was first observed by Abdel-Jawad \textit{et al.} \cite{Abdel10}, but there is no consensus on its origin. The dimer approach takes the BEDT-TTF dimer as an entity, neglecting the intra-dimer degrees of freedom; this leads to a quasi two-dimensional electron system with a half-filled conduction band, where on-site Coulomb repulsion dominates, making \CuCN ~a prime example to study the physics of genuine Mott insulators \cite{Pustogow18b}.
Alternatively, one considers a single BEDT-TTF molecule with certain inter- and intra-dimer interactions. As a consequence, \CuCN ~is regarded as a $\frac{3}{4}$-filled system, making it unstable towards a charge-ordered state, which is competing with the dimer-Mott state. Starting from the quarter-filled extended Hubbard model, some theories predict fluctuating charge disproportionation within a dimer, resulting in quantum electric dipoles \cite{Hotta10,Naka10}. The dielectric response in the audio- and radio-frequency range was interpreted as a consequence of these electric dipoles \cite{Abdel10,Hotta10,*Hotta12,Naka10,Li10,*Dayal11,*Clay12,Gomi13}; their collective optical excitations should show up in the THz region.

However, infrared and vibrational spectroscopy clearly discards a sizable charge disproportionation on the dimers \cite{Sedlmeier12}; a mixture of lattice and molecular vibrations perfectly explains all the observed optical modes even below 1~THz \cite{Dressel16}. In a recent theoretical study \cite{Fukuyama17} Fukuyama \textit{et al.} considered the crossover from a quarter-filled system with charge-ordered ground state to a dimer-Mott insulator due to strong dimerization. At high energy (in the range of eV, i.e.\ optical frequencies) the latter is stable, whereas at very low energy ($10^{-10}{\rm eV} \approx 10$~kHz) extended domains of different charge polarities arise. As a consequence, domain walls form in the system, giving rise to the HT peak. Very recently, Pouget and collaborators \cite{Foury-Leylekian18} thoroughly investigated the crystal structure of \CuCN ~and discovered a triclinic symmetry with two inequivalent dimers in the unit cell. This implies a rather weak charge imbalance between dimers in the whole temperature range.

During the last years evidence has accumulated that the interaction between the cationic (BEDT-TTF)$^{+}_{2}$ and the anionic Cu$_{2}$(CN)$_{3}^{-}$ layer is crucial for the understanding of these charge-transfer salts \cite{Dressel16,Pinteric14,Foury-Leylekian18}. The ambiguity in the arrangement of the polar CN$^{-}$ linking the triangular coordination of Cu atoms results in intrinsic disorder. Density functional theory calculations estimate that flipping a CN link, which is mainly oriented along the \textit{b}-direction, costs 174~meV, whereas flipping one that is mainly oriented along the \textit{c}-direction is only 10-15~meV \cite{Dressel16}. The interaction via hydrogen bonds maps the domains onto the BEDT-TTF layer, leading to long-range charge inhomogeneities that are detected by low-frequency probes in the kHz and MHz range \cite{Pinteric14,Pinteric15,Pinteric18}.

The extension of the relaxation time $\tau_{\rm HT}$ and decrease of $\Delta_{\rm HT}$ as pressure rises indicate that the domains increase in size and move more easily. On the other hand, a decrease of $\tau_{\rm HT}$ and barrier energy $E_{\rm VFT}$ upon x-ray irradiation, reported by Sasaki \textit{et al.} \cite{Sasaki15}, infer more domains of smaller size. This can be explained by a larger number of charged defects in the anion layer upon irradiation which act as pinning centers. The qualitatively similar but quantitatively slightly different behavior found in our sample 1 (Fig.~\ref{eps(T)_sample1}) corroborates these observations.

It is interesting to recall that also x-ray irradiation leads to a shift of the HT peak to lower temperatures \cite{Sasaki15}. High-energy irradiation produces crystal defects mainly in the anion layer and is supposed to increase the number of charge carriers. Similar to the rising pressure, the conductivity of the sample is enhanced compared to the pristine case. This clearly indicates that the relaxor-ferroelectric HT peak is influenced by screening due to free charge carriers, disentangling it from an intrinsic origin related to the conduction electrons themselves, thus rendering a scenario of intra-dimer dipoles unlikely.

\subsubsection{Low-temperature peak}

Let us come to the LT peak, which was not observed in previous ambient pressure studies of \CuCN \cite{Abdel10,Pinteric14,Sasaki15,Pinteric18}. Consistently, in both samples under inspection, we see a remarkable growth of the LT peak with pressure and a shift in its position as shown in Fig.~\ref{T_B}. The clearly distinct pressure evolution of the LT parameter compared to the HT peak  (Fig.~\ref{Fit_tau}) indicates a different origin.
As can be seen from Fig.~\ref{T_B}, the peak appears only below the critical temperature $T_{\rm crit}$, which establishes the upper bound of the coexistence regime. Note that spatially separated metallic inclusions can persist in an insulating host well before percolation sets in.
Starting from the LT peak at low $p$, the pressure evolution of $\varepsilon_{1}(p)$  upon approaching the phase boundary can be well fitted by Eq.~(\ref{eq:eps_q}) for $T<T_{\rm crit}$, as expected for a percolating system (Fig.~\ref{Fit_eps}).
Hence it is tempting to assign this feature to `metallic quantum fluctuations' previously concluded from optical spectroscopic studies \cite{Pustogow18b}.
The LT peak grows towards the phase transition and eventually becomes the dominant peak at the percolation threshold, as seen in Fig.~\ref{eps(T)}.

The energy $\Delta_{\rm LT}$ stays constant up to the IMT, where $\varepsilon_{1}(p)$ follows the BEMA model (Fig.~\ref{BEMA}) -- the latter does not consider a capacitive coupling of the metallic inclusions. Hence, we attribute the drop in $\Delta_{\rm LT}$ close to the percolation threshold to an increased coupling between the metallic inclusions.
With values of $\tau_{\rm LT}$ decreasing from $10^{-6}$ to $10^{-8}$~s, the LT relaxation is clearly slower than the HT peak; it hardens upon rising $x$. The origin of this intriguing behavior has yet to be clarified, i.e.\ whether this can be assigned to changes in size and/or shape of the metallic inclusions. Such a behavior was revealed by ellipsometric studies of VO$_2$ films \cite{Voloshenko18,*Voloshenko19}.

At this point it is worth to mention that the LT peak is reminiscent of the dielectric response in the related compound \CuCl\ \cite{Lunkenheimer12,Tomic13,Tomic15}, which is very close to the metal insulator transition already at 0~kbar.
This enables investigations of the LT peak and the nature of the metallic inclusions at ambient pressure via a broad spectrum of experimental techniques, such as scanning near-field infrared microscopy \cite{Qazilbash07}.
Deuterating $\kappa$-(BEDT\--TTF)$_2$\-Cu[N(CN)$_2$]Br crystals is another way to approach the metal-insulator transitions. In these systems Sasaki {\it et al.} succeeded to spatially map micrometer size domains using infrared spectroscopy \cite{Sasaki04}.
However, the insulating ground state is an antiferromagnet, and hence the phase boundary has opposite slope due to the Clausius-Clapeyron relation.

\subsection{Dielectric catastrophe}
The transition from insulating to metallic conduction properties manifests in numerous forms in condensed-matter physics \cite{Mott90}. Doping silicon with phosphorous, for instance, turns it metallic as the amount of electron donors exceeds a critical concentration $N_c = 3.5 \times 10^{18}~{\rm cm}^{-3}$~ \cite{ShklovskiiEfrosBook84,Lohneysen90,*Lohneysen00,*Lohneysen11}.
The transition between localized and metallic phases in disordered electronic systems is known as Anderson transitions \cite{Evers08,Abrahams10}.
Random systems or networks form long-range connectivity when crossing the percolation threshold; extended scaling theories have been developed but the details strongly depend on the particular lattice and dimension \cite{Kirkpatrick73,StaufferBook94,BollobasBook06}.
The formation of density waves due to Fermi surface nesting --~in particular in low-dimensional solids~-- leads to the opening of a gap in the density of states \cite{Gruner94,Monceau12}.
In the present case of a Mott transition, electron-electron interaction causes the metal to become insulating. The transition can be driven by either varying the electron density $N$ or the interaction strength $U$~\cite{Gebhard97,Imada98,Edwards95}.

Besides the usual thermodynamic signatures of phase transitions, the experimental hallmark of all of these insulator-to-metal transitions is a drop in resistivity, often by many orders of magnitude, upon changing the order parameter or temperature.
In addition, a divergency of the static dielectric constant is predicted by classical percolation theory
when approaching the transition from either side, with some characteristic scaling behavior
\cite{Dubrov76,Efros76,Bergman77,*Bergman78,Genzel80,Clerc90}.
Experimentally, Castner {\it et al.} first observed a strong increase of the static dielectric constant at a critical concentration $N_c$ when they measured $n$-doped silicon in the kHz range at low temperatures \cite{Castner75,*Castner79}; systematic studies of P:Si \cite{Rosenbaum83,Hering07} revealed $\varepsilon_1-\varepsilon_{\rm host} \propto (N_c/N - 1)^{1.2}$.
Most investigations deal with material mixtures, such as microemulsions \cite{Grannan81,vanDijk86,Clarkson88a,*Clarkson88b,Alexandov99},
composites \cite{Pecharroman00,*Pecharroman01,Nan10} or percolating metal films \cite{Berthier97,Hovel10,*Hovel11,deZuani14,*deZuani16}.

In many cases, the metal-insulator transition is only crossed by lowering the temperature;
here thermal fluctuations and inhomogeneities may occur. The situation is distinct from investigations of the phase coexistence by tuning the effective correlation strength via pressure in the limiting case of $T\rightarrow 0$.
Tanner and collaborators \cite{Pan76}, for instance, analyzed their temperature-dependent measurements on the charge-density-wave transition in TTF-TCNQ using a self-consistent effective-medium approximation \cite{Stroud75,*Stroud78}.
Temperature-dependent near-field infrared microscopy of Qazilbash {\it et al.} \cite{Qazilbash07,Qazilbash11,*Huffman18} on VO$_2$ films actually maps the spatial phase separation, and they extract a divergence of the dielectric constant at the transition temperature. In the case of the high-temperature Mott transition of V$_2$O$_3$, Limelette {\it et al.} concluded the coexistence region close to the critical endpoint at $T_{\rm crit} = 458$~K from the hysteresis in the pressure-dependent conductivity curves \cite{Limelette03b}.  From transport measurements on $\kappa$-(BEDT\--TTF)$_2$\-Cu$_2$[N(CN)$_2$]Cl similar conclusions were drawn on the coexistence of  Mott insulator and correlated metal \cite{Limelette03a}. In addition, NMR experiments revealed the coexistence of antiferromagnetism and superconductivity, as well as a hysteresis in susceptibility \cite{Lefebvre00}.

Considering the IMT in doped semiconductors, Sir Nevill Mott termed the divergence of the electric susceptibility `dielectric catastrophe' only in the second edition of his seminal monograph on the IMT \cite{Mott90}. Beyond percolation and polarization issues, he pointed out the importance of the localization length $\xi$ and electronic interaction. Aebischer {\it et al.} came back to this idea and theoretically analyzed the $\varepsilon_1\propto \xi^2$ behavior when approaching the Mott transition from the insulating side \cite{Aebischer01}. These situations are distinct from the present case, where phase coexistence causes the permittivity enhancement at the first-order IMT.

Record-high dielectric constants are observed when low-dimensional metals undergo a spin- or charge-density wave transition \cite{Gruner94}. Due to nesting of the Fermi surface, a gap $\Delta_0$ opens in the density of states, leading to $\varepsilon \propto \Delta_0^{-2}$ on the order of $10^6$ to $10^8$ \cite{DresselGruner02}. In general, this mechanism also holds for opening a Mott gap \cite{Aebischer01} and, therefore, we estimate the impact of reducing the gap on $\varepsilon_{1}$ in the following.

\subsubsection{Reducing the Mott gap}

\begin{figure}[h]	
	\centering \includegraphics[width=0.7\columnwidth]{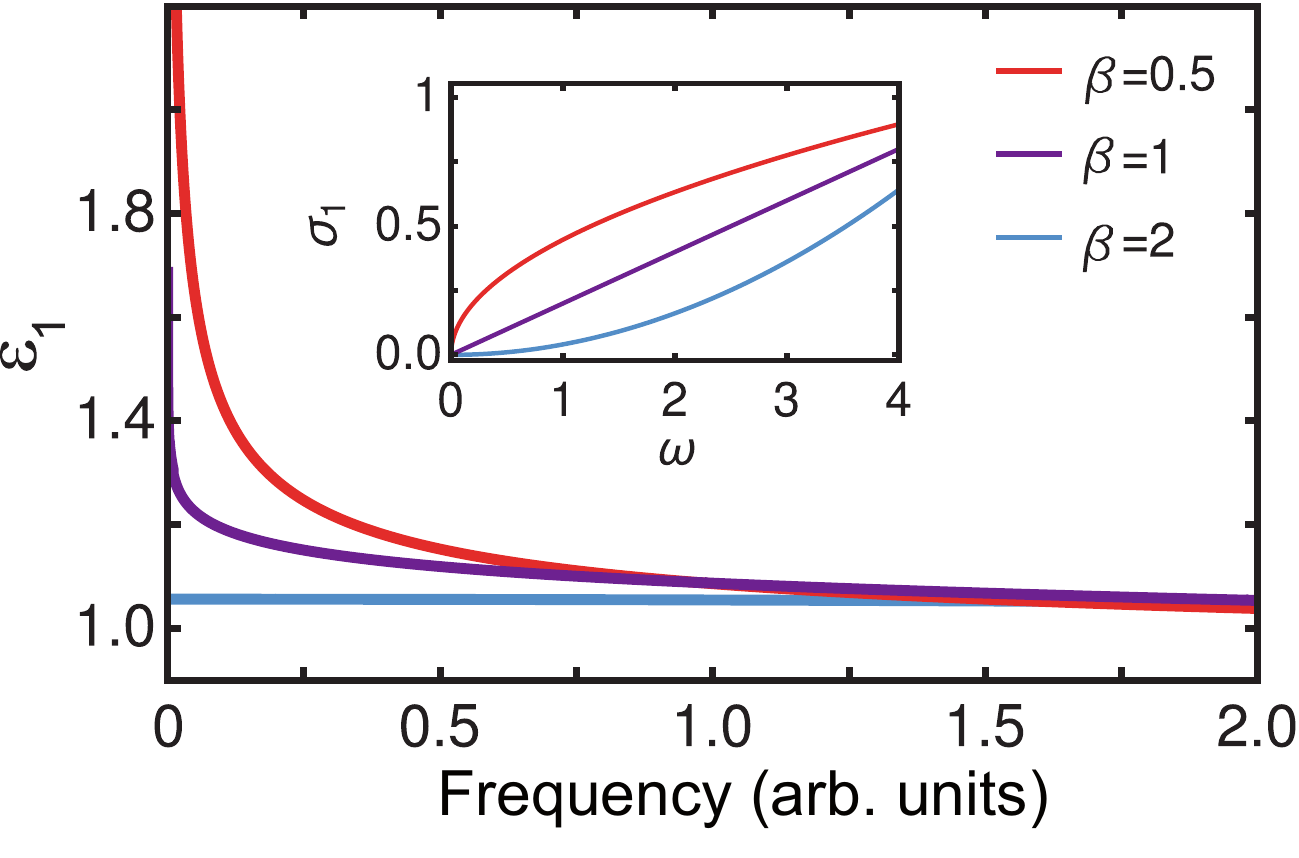}
	\caption{Frequency dependence of the real part of the dielectric constant,  $\varepsilon_{1}(\omega)$, for different power-law behavior of the optical conductivity $\sigma_1(\omega) \propto \omega^{\beta}$ as plotted in the inset.}
	\label{Theory_eps1}
\end{figure}

Let us assume that the Mott gap is already closed. Then the real part of the conductivity follows
$\sigma_{1}(\omega)\approx A\left|\omega\right|{}^{\beta}$
at low frequency, as shown in the inset of Fig.~\ref{Theory_eps1}, where $A$ is a normalization factor.
Any non-analytic behavior of the dielectric function $\varepsilon_{1}(\omega\rightarrow0)$
can only result from a sufficiently singular form of $\sigma_{1}$ also at low frequency.
The corresponding $\varepsilon_{1}(\omega)$ can be directly obtained from the Kramers-Kronig relations \cite{DresselGruner02}, yielding
\begin{equation}
\varepsilon_{1}(\omega)=1+4\int_{-\infty}^{+\infty}\frac{\sigma_{1}(\omega^{\prime})/\omega^{\prime}}{\omega^{\prime}-\omega}{\rm d}\omega^{\prime} \quad .
\end{equation}
For a linear-in-frequency increase of the optical conductivity ($\beta=1$), it is straightforward to analytically calculate
$\varepsilon_{1}(\omega)$:
\begin{equation}
\varepsilon_{1}(\omega) = -2A \log\left\{\omega\right\} + B(\omega) \quad ,
\end{equation}
where $B(\omega)$ is an analytic function that remains finite at $\omega=0$.
Fig.~\ref{Theory_eps1} displays the behavior of $\varepsilon_{1}(\omega)$ for different values $\beta$ of the power law.
For $\beta=1$, the dielectric constant $\varepsilon_{1}(\omega)$ assumes a logarithmic divergence at zero frequency,
which is a very weak singularity.
If $\beta<1,$ then $\varepsilon_{1}(\omega=0)$ diverges more strongly;
the static dielectric constant is just a finite number as $\beta$ gets larger than unity.

Within DMFT, the precise exponent $\beta$ is still under debate \cite{Eisenlohr19}, but it is at most of order unity, consistent with our numerical results. This means that the permittivity varies only little by reducing the Mott gap as the transition is approached from the insulating side and cannot cause the enormous dielectric enhancement found in our experiments.
The latter is well described by microscopic dynamical mean-field theory combined with macroscopic percolation theory \cite{Pustogow2021-percolation}.
In the present case of \CuCN, optical investigations at ambient pressure unambiguously show that no clear-cut gap exists, despite the strongly insulating behavior \cite{Kezsmarki06,Elsaesser12}. In the THz and far-infrared range the conductivity is rather well described by a linear frequency dependence, corresponding to $\beta \approx 1$ \cite{Dressel18}.
The low-energy spectral weight actually increases upon lowering the temperature, in contrast to the quantum spin liquid  compounds \AgCN\ and $\beta^{\prime}$-EtMe$_3$\-Sb\-[Pd(dmit)$_2$]$_2$, which are all far away from the IMT boundary \cite{Pustogow18b}.
Low-temperature optical investigations on $\kappa$-[(BEDT\--TTF)$_{1-x}$-(BEDT\--STF)$_{x}$]$_2$\-Cu$_2$(CN)$_3$ in the far-infrared range yield that around 1~THz $\varepsilon_1(x)$ increases only slightly when spectral weight is transferred to low frequency as the IMT is approached by substitution (see Supplementary Fig. 7 in Ref. \cite{Pustogow2021-percolation}).
Hence, we conclude that the observed enhancement of the permittivity upon approaching the Mott transition originates from
the first-order nature and the concomitant percolating phase coexistence, and is not caused by closing the Mott gap.

\subsubsection{Phase coexistence at the genuine Mott transition in organic spin liquids}

The enhancement of $\varepsilon_1(U/W)$ --~respectively $\varepsilon_1(p)$~-- is sharply confined to the coexistence region, where it exceeds the values of the (homogeneous) Mott insulator by orders of magnitude.
In related studies on $\kappa$-[(BEDT\--TTF)$_{1-x}$-(BEDT\--STF)$_{x}$]$_2$\-Cu$_2$(CN)$_3$
chemical substitution was utilized to increase the bandwidth in order to tune the system across the Mott transition \cite{Pustogow2021-percolation,Pustogow2021-Landau}.
The dielectric permittivity exhibits a similar maximum when approaching the phase boundary as correlations decrease. In the coexistence regime, percolation and correlation effects both contribute to the dielectric properties. It is interesting to note that this peak appears in an abrupt fashion, very distinct from the smooth and gradual increase in disorder-driven IMTs \cite{Dyre00}.
Moreover, around the first-order IMT below $T_{\rm crit}$, the permittivity exhibits a pronounced frequency dependence beyond standard percolation theory \cite{Pustogow2021-percolation}.

Finally, we draw attention to recent DMFT calculations \cite{Eisenlohr19} revealing a crossover (non-asymptotic) power-law behavior in the spectral function $A(\omega)$ and the self-energy $-{\rm Im}\left\{\Gamma(\omega)\right\}$ extending from low to elevated temperatures. Concomitantly, consistent scaling of the resistivity is found above and below $T_{\rm crit}$ as well. Both are traced back to the metastable insulating phase in the coexistence region, suggesting local quantum criticality of the Mott transition below $T_{\rm crit}$, which eventually is also responsible for its well-known counterpart at elevated temperatures. Besides these fundamental findings, this study reveals a peculiar low-frequency behavior of $A(\omega)$ in proximity to the Mott transition. Whether this can be connected to the intriguing dielectric response observed here has yet to be clarified, but might provide a route to place the various features in $\varepsilon_{1}(p,T,\omega)$ on the same footing, such as the pressure evolution of the HT peak and the anomalous power-law decrease of $\varepsilon_{1}(\omega)$ in the coexistence region. We remind at this point that the HT peak is observed in several other organic dimer Mott insulators \cite{Abdel13,Iguchi13,Pinteric16} with triangular lattice, showing that its emergence is independent of details in the crystal structure.

\subsubsection{Applicability of percolation theory}

Percolation theory is applicable for systems consisting of two
distinct types of domains. One should keep in mind, however, that  in real materials domain walls
are always present with properties distinct from either of the coexisting phases.
The precise characteristics of the domain walls reflect the specifics of the clean system displaying
phase coexistence and, as such, have particular dimensions (thickness) and occupy
a finite volume fraction. The effects of such domain walls can be expected to be negligible
if their dimension (thickness) is much smaller than the characteristic domain size, which
is what we expect for weak disorder. In contrast, when disorder is sufficiently strong,
it is expected \cite{Imry75,DobrosavljevicBook12} to produce nucleation centers for more and more droplets,
leading to the
reduction of the domain size, which eventually becomes comparable to the thickness of
the domain walls. When this happens, a simplistic two-component percolation picture is no
longer of direct relevance, and one may expect more gradual variations of all observables compared to simple percolation theory. We believe that finite disorder is the
main reason why the peak of $\epsilon_{1}(p)$ appears more narrow in theoretical results than in experiments (cf.\ Fig.~4 in \cite{Pustogow2021-percolation}).

We want to recall that in contrast to most other examples of percolative behavior, here we do not have different materials mixed, not even the crystal structure or symmetry changes between insulating and metallic regions. In the ideal case, the domains are distinct by the effect of correlations on their physical properties.
From thermal expansion studies on the related compound \CuCl\ in the vicinity of the critical endpoint, we know that metallic and
insulating phases exhibit a slightly different volume and distinct expansion coefficients \cite{Gati16}.
The particular arrangement is susceptible to strain, impurities, etc. leading to domain boundaries with an intermediate lattice constant on a local scale.

\subsection{Phase diagram}
The main results of our pressure- and temperature-dependent dielectric spectroscopic studies on \CuCN\ are summarized in Figs.~\ref{Phase_diagram} and~\ref{theory}.
The former displays a three-dimensional plot of $\varepsilon_{1}(p,T)$ measured at $f=380$~kHz. The bottom area contains a sketch of the phase diagram constructed on the projection of the $\varepsilon_{1}(p,T)$ values with the corresponding color code; the intense dark red area indicates the enhanced values in the coexistence phase when spatially separated metallic regions grow in the insulating matrix. The percolative behavior softens as temperature increases: the maximum diminishes and eventually a gradual crossover remains above $T_{\rm crit}$. Additionally, we include the quantum Widom line from the data of Ref. \onlinecite{Furukawa15} that agrees with the results presented here. The bifurcation temperature $T_{\rm B}$ marks the change from the HT to the LT peak and the concomitant modification in the relaxation dynamics.
The Fermi-liquid temperature $T_{\rm FL}$ was extracted from the resistivity \cite{Pustogow2021-percolation}; our findings are in accordance with previous reports \cite{Kurosaki05,Furukawa18}.

Fig.~\ref{theory} displays the phase diagram of a genuine Mott insulator around the first-order phase transition to the metallic state.
For each of the three ranges we compare experimental and theoretical results of the conductivity and dielectri permittivity.
Our data provide first experimental evidence for the coexistence of the Mott-insulating and the metallic phases, as predicted by DMFT calculations on a disordered Hubbard model
for half-filling \cite{Byczuk05,Vucicevic13}. This regime of metal-insulator coexistence emerges from the insulating phase and partially overlaps with the Fermi-liquid regime. When the phase boundary is crossed, the metallic fraction grows and quickly forms a continuous path through the specimen; the capacitive coupling of remaining metallic puddles in the insulating regions leads to the large values of $\varepsilon_{1}$. The percolative behavior is strongly suppressed for higher temperatures where the first-order Mott IMT becomes a smooth crossover and the contrast in conductivity between metallic and insulating fraction diminishes.

Approaching the phase boundary from the insulating side, we determine a critical exponent $q \approx 1$ in Eq.~(\ref{eq:eps_q}) that is consistent with Bruggeman's effective medium approximation that allowed us to extract the metallic volume filling fraction as a function of pressure.
%The frequency dependence $\varepsilon_{1}(\omega)$ determined right at the percolation threshold ($p=1.7$~kbar) indicates that tunneling between adjacent metallic clusters \cite{Pakhomov98,Sarychev94} takes place over distances up to $d_{t} \approx 23$~\AA.
%Finally, our findings are consistent with scanning infrared micro reflectance spectroscopy (SIMS) measurements on partially deuterated $\kappa$-($h8$-BEDT-TTF)$_{1-x}$($d8$-BEDT-TTF)$_x$Cu[N(CN)$_2$]Br, located at the metal insulator transition, which provides evidence for a phase separation into metallic and insulating domains with 50-100~$\mu$m in diameter \cite{Sasaki04}.

\section{Summary}
\label{Summary}
Our dielectric measurements as a function of frequency, temperature and pressure in the insulating state of \CuCN\ reveal that the relaxor-ferroelectric peak below $T=50$~K shifts to lower temperatures as pressure increases because the screening by free charge carriers becomes pronounced with increasing bandwidth. A second peak emerges at lower temperatures and grows in amplitude with applying pressure. We attribute this behavior to the sparse occurrence of metallic puddles in the insulating host phase.
Upon moving deeper into the phase coexistence region of the first-order transition from the Mott-insulating to the metallic phase,
we discover a strong enhancement of $\varepsilon_{1}(p)$ up to $10^{5}$ at lowest frequencies ($f=7.5$~kHz), which resembles percolating behavior. We apply Bruggeman's effective medium approximation to determine the metallic filling fraction and obtain a critical exponent $q\approx 1$ upon approaching the threshold $x_c$  from the insulating side:
$\varepsilon_{1}(x) \propto (x_{c} - x)^{-q}$. Calculations by dynamical-mean-field theory on a single-band Hubbard model reproduce our comprehensive experimental findings in full breadth. The divergency of the dielectric permittivity is mainly caused by classical percolation physics of a strongly correlated electron system close to the Mott transition.
Our results provide compelling evidence for the coexistence of metallic and insulating regions, and we demonstrate the capabilities of dielectric spectroscopy as a `smoking gun' to probe phase coexistence and spatial inhomogeneities at metal-insulator transitions.

\section{acknowledgments}
We thank G. Untereiner for contacting the crystals. We acknowledge support by the Deutsche Forschungsgemeinschaft (DFG) via DR228/52-1. A.P. acknowledges support by the Alexander von Humboldt-Foundation through the Feodor Lynen Fellowship. E.U. acknowledges the support of the European Social Fund and by the Ministry of Science Research and the Arts of Baden-W{\"u}rttemberg.
Work in Florida was supported by the NSF Grant No. 1822258, and the National High Magnetic Field Laboratory through the NSF Cooperative Agreement No. 1644779 and the State of Florida.

%\clearpage

\appendix

\section{Dielectric properties of sample 1}
\label{sec:sample1}

\begin{figure}
	\centering
	\includegraphics[width=1.0\columnwidth]{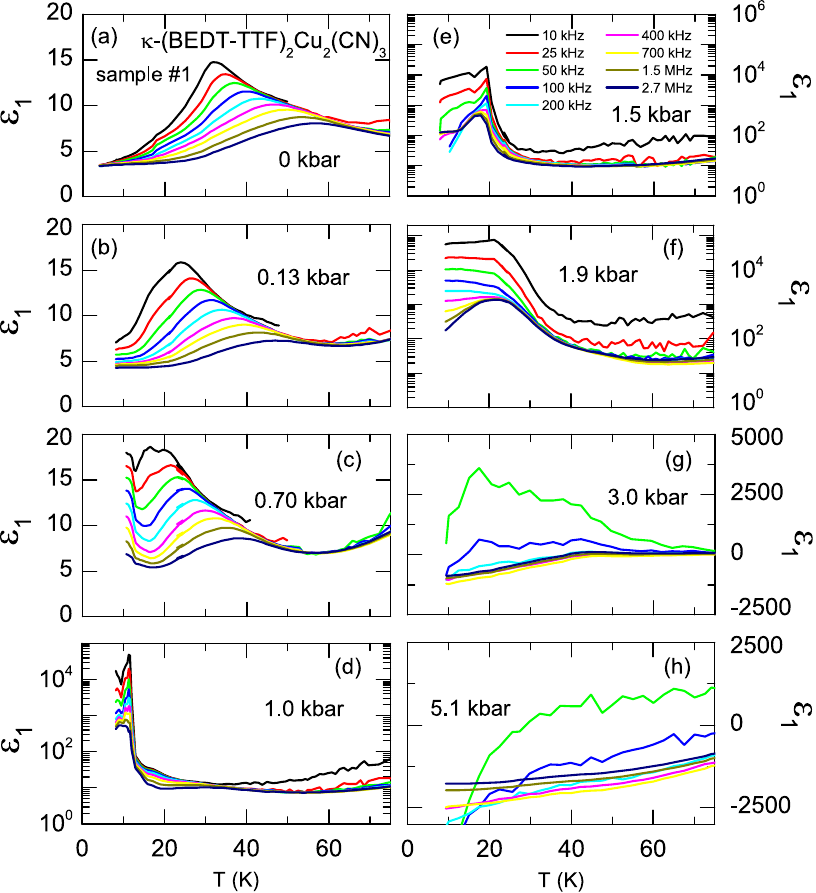}
	\caption{Plot of the dielectric permittivity $\varepsilon_{1}(T)$ of \CuCN\ sample 1 for several frequencies upon increasing pressure. The results are qualitatively identical to those obtained on sample 2, displayed in Figs.~\ref{eps(T)} and~\ref{eps(T-high-p)}. (a) At ambient pressure the relaxor-type ferroelectric peak starts already around $T=70$~K and reaches the maximum at low frequencies at 30~K.
(b-c)~A shoulder-like feature is already present at ambient pressure and develops into a second peak upon pressurizing. (d-f) In the coexistence phase between $p=1.0$ and 1.9~kbar, an enormous increase of $\varepsilon_{1}$ is observed due to percolation. (g,h) Above $p=3.0$~kbar, $\varepsilon_{1}<0$ for nearly all measured frequencies indicating metallic behavior.}
	\label{eps(T)_sample1}
\end{figure}

Most data presented in this article have been obtained from dielectric measurements on sample 2; the conclusions on phase separation at the Mott IMT are fully supported by similar findings on sample 1. Fig.~\ref{eps(T)_sample1} gives an overview on the pressure evolution of its dielectric response by plotting $\varepsilon_{1}(T)$ for selected frequencies and pressures as indicated. The position of the HT feature at ambient pressure, for instance at 40~K when probed at $f=100$~kHz, is in good agreement with previous ambient pressure reports \cite{Abdel10,Pinteric15}. Unfortunately, only a few measurements in the Mott-insulating phase could be performed before the sample broke. Nevertheless, despite the low pressure resolution, it is obvious that the shift of the HT peak upon increasing pressure resembles the findings for sample 2. In contrast, the LT peak for sample 1 is barely visible up to $p=0.45$~kbar and becomes evident only at 0.7~kbar around $T=10$~K. The strong enhancement of $\varepsilon_{1}$ by several orders of magnitude upon entering the coexistence regime is already observed at $p=1.0$~kbar and extends up to 1.9~kbar [Fig.~\ref{eps(T)_sample1}(f)]. We emphasize that this is in accordance with the shift of the IMT in sample 1. Interestingly, the plateau-like shape of $\varepsilon_{1}(T)$ is observed only for $f \leq 400$~kHz whereas for higher frequencies a slight downturn occurs. Upon further pressure increase, metallic behavior with $\varepsilon_{1}<0$ sets in.% whereas only for the lowest frequencies positive values for the permittivity remain.

\begin{figure}
	\centering
	\includegraphics[width=0.8\columnwidth]{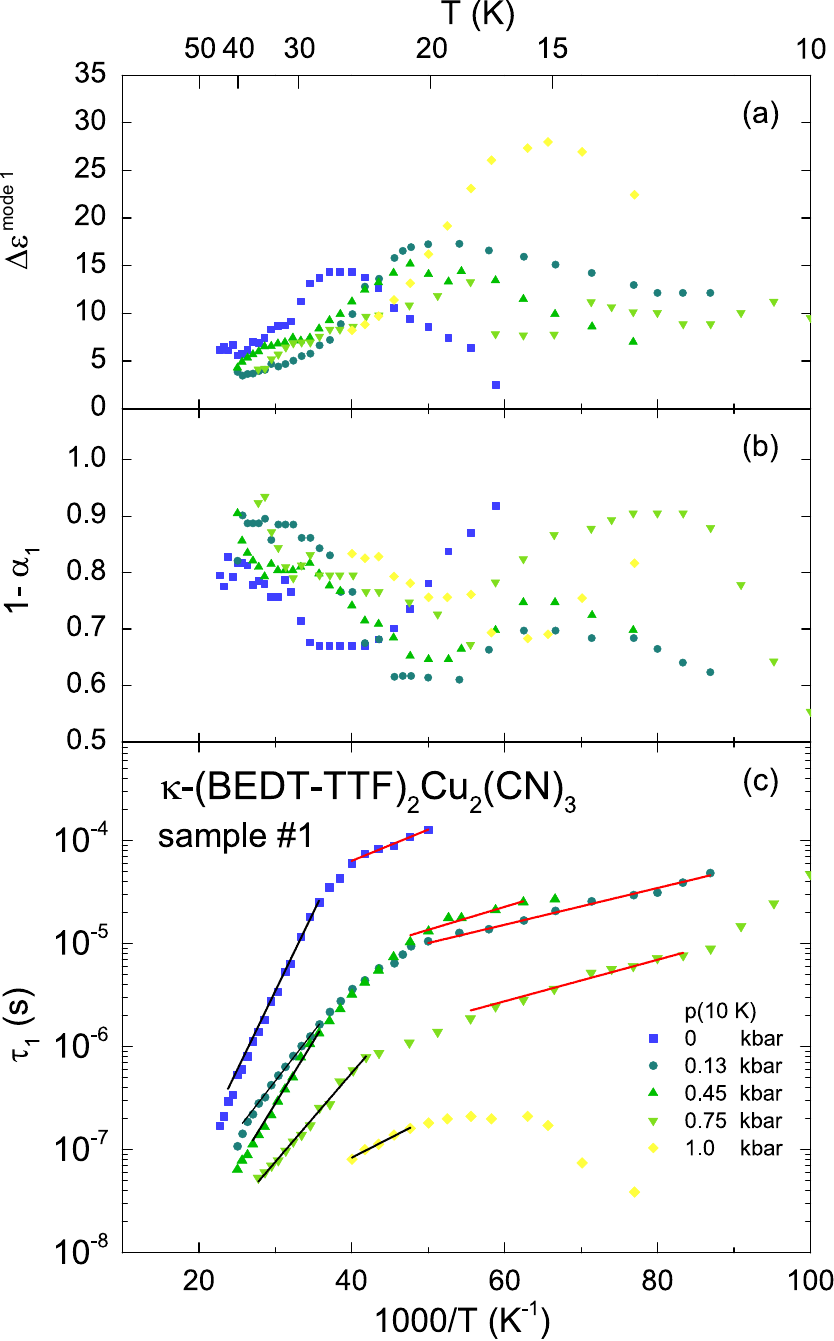}
	\caption{Arrhenius plot of the fitting parameters of mode 1 for sample 1 at different pressures as indicated. (a)~Dielectric strength $\Delta \varepsilon^{\text{mode 1}}(T)$, (b)~distribution of relaxation times $1-\alpha_{1}(T)$ and (c)~mean relaxation time $\tau_{1}(T)$. The black and red lines represent fits with Eq.~(\ref{eq:HT}) above and below the kink in $\tau_{1}(T)$ at $T_{\rm{B}}$, respectively.}
	\label{Debye_results_mode1_sample1}
\end{figure}

\begin{figure}
	\centering
	\includegraphics[width=0.8\columnwidth]{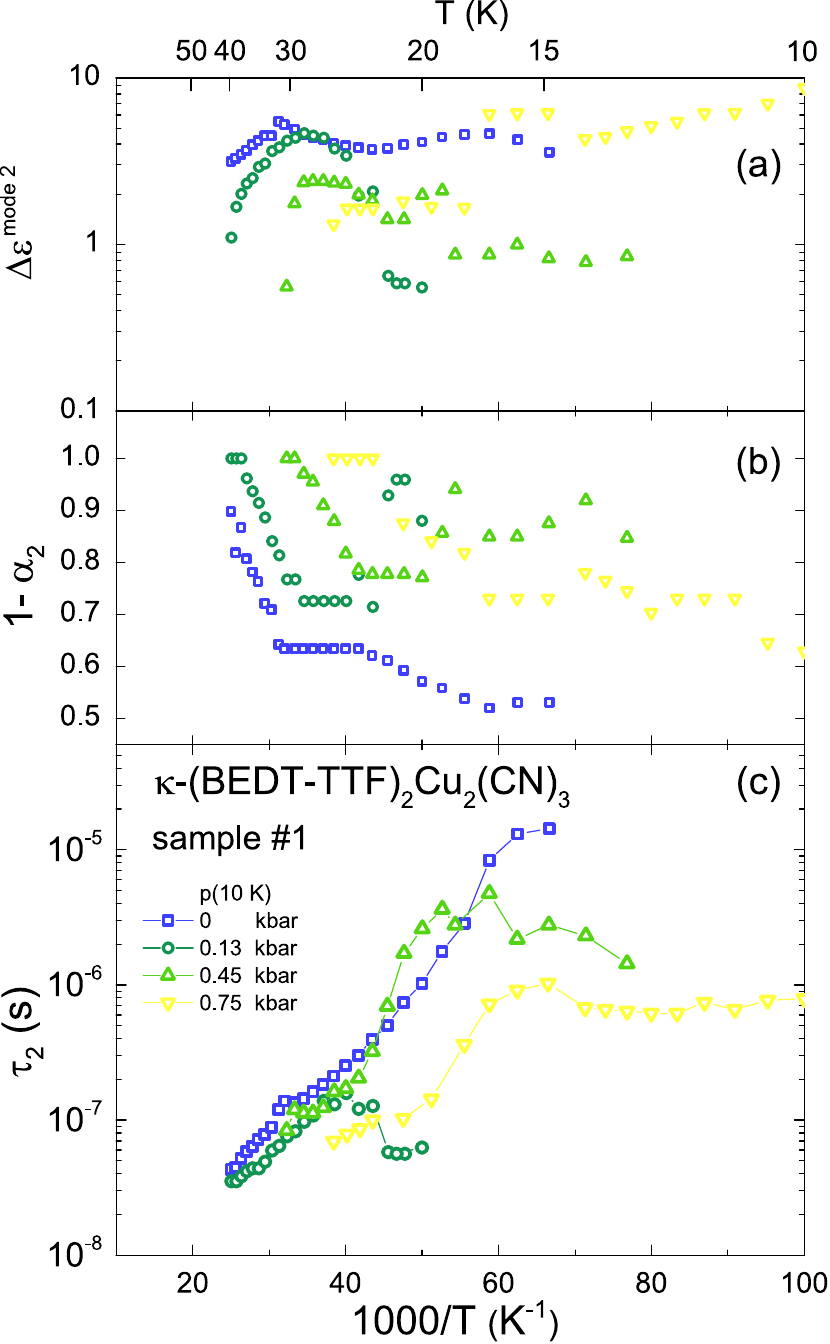}
	\caption{Arrhenius plot of the fit parameters of mode 2 for \CuCN\ sample 1 at different pressures as indicated. (a)~Dielectric strength $\Delta \varepsilon^{\text{mode 2}}(T)$, (b)~distribution of relaxation times $1-\alpha_{2}(T)$, and (c)~mean relaxation time $\tau_{2}(T)$.}
	\label{Debye_results_mode2_sample1}
\end{figure}

In order to analyze the frequency-dependent permittivity of these data, we have applied the same procedure as presented in Section~{sec:insulator} on sample 2. The obtained fit parameters  $\Delta \varepsilon^{\text{mode 1}}$, $1-\alpha_{1}$ and $\tau_{1}$ for mode 1 are plotted in Fig.~\ref{Debye_results_mode1_sample1} as a function of inverse temperature. The peak in $\Delta \varepsilon^{\text{mode 1}}(T)$ shifts to lower $T$ and increases in amplitude as pressure is applied. This behavior is reminiscent to what is observed for sample 2, however, it is less pronounced for sample 1, for which the LT peak is weaker and becomes apparent only at $p\geq 1.0$~kbar. This also explains the higher values of $1-\alpha_{1}$ and their restrengthening upon cooling, the latter most pronounced at ambient pressure, which indicate less broadening in sample 1. In $\tau_{1}(T)$, a kink at $T_{\rm{B}}$ is observed, which shifts to lower temperatures while the corresponding relaxation time gets shorter, reproducing the behavior observed in sample 2. The parameters obtained by fitting $\tau_{1}(T)$ with the activated behavior [Eq.~(\ref{eq:HT})] for the HT mode ($T>T_{\rm{B}}$) and the LT mode ($T<T_{\rm{B}}$) are already presented and discussed in Table~\ref{tab:Fit_Delta} and  Fig.~\ref{Fit_tau} above.

Fig.~\ref{Debye_results_mode2_sample1} displays the parameters $\Delta \varepsilon^{\text{mode 2}}(T)$, $1-\alpha_{2}(T)$ and $\tau_{2}(T)$ of mode 2 for sample 1 of \CuCN. The dielectric strength $\Delta \varepsilon^{\text{mode 2}}$ exhibits a maximum around $T=30$~K at ambient pressure, which shifts towards lower temperatures and diminishes upon increasing $p$. This is in contrast to sample 2, for which $\Delta \varepsilon^{\text{mode 2}}$ monotonously grows upon cooling (cf.\ Fig.~\ref{Debye2}). Throughout the entire pressure range $\Delta \varepsilon^{\text{mode 2}}$ is smaller than the one of mode~1 by approximately a factor of 2. The mode shifts towards lower temperatures with rising pressure. A drop in $1-\alpha_{2}(T)$ indicates a considerable broadening of the mode upon cooling, which gets less pronounced for increasing pressure. The temperature dependence of $\tau_{2}$ is strongly influenced by pressure and shows sample dependence. For sample 1, the monotonic increase is much steeper and observed down to 15~K at ambient pressure which saturates into a plateau upon increasing pressure. In contrast to sample 2, only the onset of a shallow minimum around $T=12$~K is revealed indicating non-monotonic relaxation dynamics.

\section{Spurious effects}

\subsection{Contacts}
\begin{figure}
	\centering
	\includegraphics[width=1.0\columnwidth]{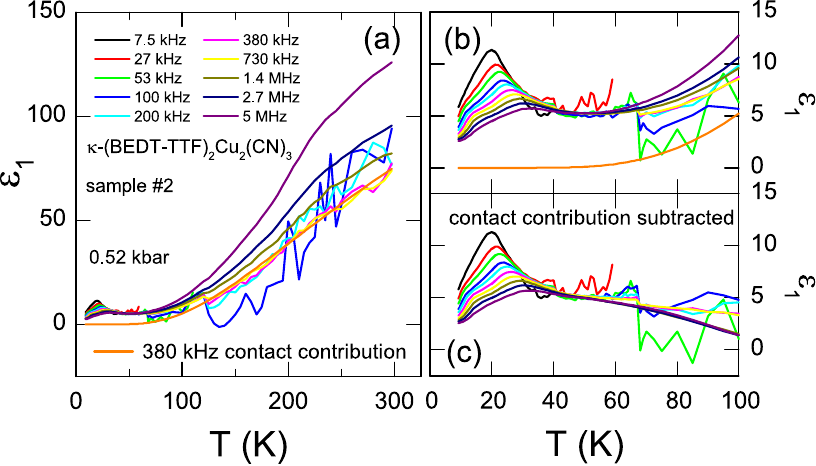}
	\caption{(a) Temperature dependence of $\varepsilon_{1}(T)$ plotted in the entire temperature range, for an applied pressure of $p = 0.52$~kbar measured at various frequencies $f$. In addition, we determine the contact contribution by fitting the high-temperature part with Eq.~(\ref{eq:Schottky_C_depl}), as shown for the example of $f=380~$kHz (orange line). (b)~Detailed view of the relaxor-ferroelectric relaxation at low temperatures including the contact contribution. (c)~Relaxor-ferroelectric response after the contact contribution has been subtracted: the contact contribution is negligible below $T=60$~K.}
	\label{Demo_subtract_contacts}
\end{figure}

Owing to the two-point configuration usually applied in dielectric spectroscopy, the obtained data may include contributions from polarization effects at the contacts, which have to be conscientiously ruled out or determined.
Since the contacts are produced by amorphous carbon (carbon paste) with metallic properties,
Schottky contacts may form at the sample-contact interfaces, resulting in a depletion layer at the interface with thickness
\begin{equation}
d_{\textrm{depl}} \ = \ \left[\frac{2 \varepsilon_{1} \varepsilon_{0}}{e N_{\textrm{c}}} (\Phi_{\textrm{m}}-\Phi_{\textrm{s}} \pm e U)\right]^{1/2} \quad , \label{eq:Schottky_depl}
\end{equation}
wherein $\Phi_{\textrm{m}}$ and $\Phi_{\textrm{s}}$ the distances between the vacuum level and the chemical potential of the metal and semiconductor, respectively, and $U$ indicates the voltage of the applied ac signal. The charge-carrier density
$N_{\textrm{c}}(T) \propto \exp\left\{- \Delta/k_{\textrm{B}}T\right\}$
is determined by thermal excitations across the charge gap $\Delta$.
The modified charge density in the depletion zone gives rise to an additional capacitance
\begin{equation}
C_{\textrm{depl}} \  \propto \ \frac{1}{d_{\textrm{n}}} \ \propto \ C  \exp \left\lbrace \frac{-\Delta}{2k_{\textrm{B}}T} \right\rbrace  \quad , \label{eq:Schottky_C_depl}
\end{equation}
which eventually is responsible for the spurious effects attributed to the contact contribution. On the right side of Eq.~(\ref{eq:Schottky_C_depl}) we estimated the temperature dependence of the contact contribution, which is governed by $N_{\textrm{c}}(T)$. In Fig.~\ref{Demo_subtract_contacts}(a) we exemplarily plot $\varepsilon_{1}(T)$ up to room temperature, probed at various frequencies as indicated. In the temperature range  $75~{\rm K} < T < 300$~K, we observe a decrease of $\varepsilon_{1}(T)$ upon cooling which is very well described by Eq.~(\ref{eq:Schottky_C_depl}) (orange line) and hence is attributed to the contact contribution.

Most importantly, the contact contribution is negligible below 60~K [Fig.~\ref{Demo_subtract_contacts}(b)] and does not influence the analysis of the relaxor like dielectric response [Fig.~\ref{Demo_subtract_contacts}(c)]. We also note, that this effect would be too weak to explain the huge enhancement of $\varepsilon_{1}(T)$ close to the phase boundary. If we inspect $\rho_{1}(T)$ and $\varepsilon_{1}(T)$ at 1.91~kbar \cite{Pustogow2021-percolation}, for instance, we find that $\rho_{1}(p={1.91~\text{kbar}},T=10~\text{K})\approx \rho_{1}(p={1.91~\text{kbar}},T=300~\text{K})$, such that the spurious contact contribution to $\varepsilon_{1}$
should be the same at 10~K and 300~K. On the other hand, we see that $\varepsilon_{1}(p={1.91~\text{kbar}},T=10~\text{K}) \gg \varepsilon_{1}(p={1.91~\text{kbar}},T=300~\text{K})$ (cf.\ Fig.~\ref{eps(T)}), which cannot be explained by
such a contact contribution, corroborating our phase coexistence scenario.

\clearpage
%\bibliographystyle{apsrev4-1}
%\bibliography{References_PRB}
%merlin.mbs apsrev4-1.bst 2010-07-25 4.21a (PWD, AO, DPC) hacked
%Control: key (0)
%Control: author (72) initials jnrlst
%Control: editor formatted (1) identically to author
%Control: production of article title (-1) disabled
%Control: page (0) single
%Control: year (1) truncated
%Control: production of eprint (0) enabled
%
\end{document}